\documentclass[12pt,a4paper]{article}

\usepackage[a4paper,vmargin=2.5cm,hmargin=2.1cm]{geometry}
\usepackage{lmodern}
\usepackage[T1]{fontenc}
\usepackage{amsmath,amssymb}
\usepackage[bookmarks=true,hyperfigures=true]{hyperref}
\usepackage{graphicx} 
\usepackage[font=small,labelfont=bf,width=0.85\textwidth]{caption} 
\usepackage[nosort]{cite} 
\usepackage{dsfont}
\usepackage[bulletsep]{collref} 
\usepackage{fixmath} 
\usepackage{tikz}
\usepackage{array,multirow}

\usepackage{fancybox}

\graphicspath{{./figures/}}

\makeatletter \let\@keywords\@empty \let\@subject\@empty
\providecommand{\keywords}[1]{\gdef\@keywords{#1}}
\providecommand{\subject}[1]{\gdef\@subject{#1}}
\def\thetitle{\@title}
\def\theauthor{\@author}
\def\thesubject{\@subject}
\def\thedate{\@date}
\def\thekeywords{\@keywords}
\makeatother
\AtBeginDocument{%
\hypersetup{pdftitle={\thetitle}}%
\hypersetup{pdfauthor={\theauthor}}%
\hypersetup{pdfsubject={\thesubject}}%
\hypersetup{pdfkeywords={\thekeywords}}%
}

\providecommand{\hypersetup}[1]{}
\providecommand{\texorpdfstring}[2]{#1}

\hypersetup{plainpages=false}
\hypersetup{pdfpagemode=UseNone}
\hypersetup{bookmarksnumbered=true}
\hypersetup{pdfstartview=FitH} 
\hypersetup{colorlinks=false}
\hypersetup{citebordercolor={.5 1 .5}}
\hypersetup{urlbordercolor={.5 1 1}}
\hypersetup{linkbordercolor={1 .7 .7}}

\numberwithin{equation}{section}

\let\oldbfseries=\bfseries
\let\oldmdseries=\mdseries
\let\oldnormalfont=\normalfont
\renewcommand{\bfseries}{\oldbfseries\boldmath}
\renewcommand{\mdseries}{\oldmdseries\unboldmath}
\renewcommand{\normalfont}{\oldnormalfont\unboldmath}

\newcommand{\sfrac}[2]{{\textstyle\frac{#1}{#2}}}

\ifx\genfrac\sdflkaj
\newcommand{\atopfrac}[2]{{{#1}\above0pt{#2}}}
\else
\newcommand{\atopfrac}[2]{\genfrac{}{}{0pt}{}{#1}{#2}}
\fi

\newcommand{\gen}[1]{\mathfrak{#1}}

\newcommand{\alg}[1]{\mathfrak{#1}}

\newcommand{\superN}{\mathcal{N}}
\newcommand{\sym}{$\superN=4$ SYM}

\newcommand{\charge}{Q}

\newcommand{\idop}{\mathds{1}}
\newcommand{\permop}{\mathds{P}}
\newcommand{\uR}{\mathcal{R}}
\newcommand{\Smat}{\mathrm{S}}

\newcommand{\braiding}{\mathcal{U}}

\newcommand{\levz}{\mathfrak{J}}
\newcommand{\levo}{\mathfrak{\widehat J}}

\newcommand{\markarrow}[2]{{\arraycolsep0pt\begin{array}[b]{c}\makebox[0cm]{$\atopfrac{#2}{\downarrow}$}\\#1\end{array}}}

\newcommand{\comm}[2]{[#1,#2]}

\newcommand{\ket}[1]{|#1\rangle}

\newcommand{\braket}[1]{\langle #1 \rangle}

\makeatletter
\def\mr@ignsp#1 {\ifx\:#1\@empty\else #1\expandafter\mr@ignsp\fi}%
\newcommand{\multiref}[1]{\begingroup
\xdef\mr@no@sparg{\expandafter\mr@ignsp#1 \: }%
\def\mr@comma{}%
\@for\mr@refs:=\mr@no@sparg\do{\mr@comma\def\mr@comma{,}\ref{\mr@refs}}%
\endgroup}
\makeatother

\newcommand{\hypref}[2]{\ifx\href\asklfhas #2\else\href{#1}{#2}\fi}
\newcommand{\secref}[1]{Section~\multiref{#1}}

\newcommand{\tabref}[1]{Table~\multiref{#1}}

\renewcommand{\eqref}[1]{(\multiref{#1})}

\makeatletter
\newlength{\apb@width}
\newcommand{\autoparbox}[2][c]{\settowidth{\apb@width}{#2}\parbox[#1]{\apb@width}{#2}}
\newcommand{\includegraphicsbox}[2][]{\autoparbox{\includegraphics[#1]{#2}}}
\makeatother

\title{Nonlocal Symmetries and Factorized Scattering}

\author{%
Florian Loebbert,
Anne Spiering
}

\begin{document}

\pdfbookmark[1]{Title Page}{title}

\thispagestyle{empty}

\begin{flushright}
\texttt{TCDMATH 18-07}

\texttt{HU-EP-18/17}
\end{flushright}

\vspace{1cm}

\begin{center}

\begingroup\Large\bfseries\thetitle\par\endgroup

\vspace{1cm}

\begingroup\scshape
Florian Loebbert$^a$ and Anne Spiering$^{a,b}$ 
\endgroup

\vspace{1cm}

\begingroup\itshape
$^a$Institut f\"ur Physik and IRIS Adlershof,\\ 
Humboldt-Universit\"at zu Berlin, \\
Zum Gro{\ss}en Windkanal 6, 12489 Berlin, Germany
\vspace{3mm}

$^b$School of Mathematics \& Hamilton Mathematics Institute\\
Trinity College Dublin\\
Dublin, Ireland
\par\endgroup

\bigskip

{\ttfamily
\href{mailto:loebbert@physik.hu-berlin.de}{loebbert@physik.hu-berlin.de}\\
\href{mailto:spiering@maths.tcd.ie}{spiering@maths.tcd.ie}
}

\vspace{3cm}

\textbf{Abstract}

\bigskip

\begin{minipage}{12.4cm}
Conventionally, factorized scattering in two dimensions is argued to be a consequence of the conservation of local higher charges. However, integrability may well be realized via nonlocal charges, while higher local charges are not known.
Here we address the question of whether a nonlocal Yangian symmetry implies factorized scattering of the S-matrix.  We explicitly study the constraints on three-particle scattering processes of particles transforming in the fundamental representations of $\alg{su}(N)$, $\alg{u}(1|1)$, and the centrally extended $\alg{su}(2|2)$ underlying the dynamic scattering and hexagon form factors in AdS/CFT. 
These considerations shed light on the role of the Yangian as an axiomatic input for the bootstrap program for integrable theories.
\end{minipage}

\end{center}

\vfill



\newpage

\setcounter{tocdepth}{2}
\hrule height 0.75pt
\pdfbookmark[1]{\contentsname}{contents}
\tableofcontents
\vspace{0.8cm}
\hrule height 0.75pt
\vspace{1cm}

\section{Introduction}
\label{sec:intro}

The most intuitive definition of integrability relies on the existence of a tower of local
 conserved charges which outnumber the degrees of freedom of a physical model or process.
It is well known that higher \emph{local}%
\footnote{Here by `local' we mean that the action of the charges on multi-particle states is realized via the trivial tensor product action.} 
symmetries imply that the scattering matrix is trivial in $d>2$ spacetime dimensions. This is the content of the Coleman--Mandula theorem. Moreover, for field theories in $d=2$ dimensions, higher local charges imply factorized scattering \cite{Kulish:1975ba,Shankar:1977cm,Iagolnitzer:1977sw,Iagolnitzer:1978my,Parke:1980ki}. In fact, in many cases two local charges with different Lorentz spin are sufficient for factorization~\cite{Parke:1980ki}. These observations generalize to other integrable models with a notion of scattering, like spin chains.

On the other hand, the same factorization arguments do not apply to \emph{nonlocal} charges (e.g.\ Yangian symmetry \cite{Bernard:1992ya,MacKay:2004tc,Torrielli:2011gg,Loebbert:2016cdm}), which furnish the mathematical underpinnings of many integrable models (e.g.\ the XXX spin chain, the Gross--Neveu model or the AdS/CFT duality). While nonlocal symmetries often coexist with higher local charges (see e.g.\ \cite{MacKay:1998kx}), it is generically not clear whether nonlocal charges imply the existence of local charges. Note for instance that only for certain models local and nonlocal charges can be constructed from the same monodromy matrix, cf.\ \cite{Tarasov:1983cj}.

In fact, in $d>2$ there are examples of scattering amplitudes which are invariant under nonlocal charges, while a local formulation of this symmetry is not known: In $d=4$, the S-matrix of $\superN=4$ SYM theory (at least at tree level) \cite{Drummond:2009fd} as well as the S-matrix of the recently found fishnet theories \cite{Chicherin:2017cns,Chicherin:2017frs} are Yangian invariant. In $d=3$ the tree-level scattering in $\superN=6$ superconformal Chern--Simons (alias ABJM) theory provides a valid case \cite{Bargheer:2010hn}.%
\footnote{Note that the Yangian enters these higher dimensional theories via the AdS/CFT duality and the underlying mechanisms are far from being fully understood. Effectively, also here the Yangian operates in one dimension, i.e.\ on the one-dimensional color trace.}
This yields a natural question: What are the implications of nonlocal symmetries on scattering processes in physical models?

\begin{table}
\begin{center}
\begin{tabular}{|l||c|c|}\hline
&$d=2$&$d>2$\\\hline\hline
\rule{0pt}{28pt}
Local charges& \begin{minipage}{5cm}\centering factorized S-matrix\par [Kulish, Shankar--Witten, Iagolnitzer, Parke, \dots]\end{minipage}&\begin{minipage}{7cm}\centering trivial S-matrix \par [Coleman--Mandula]\end{minipage}
\\\hline
\rule{0pt}{22pt}
Nonlocal charges&\begin{minipage}{5cm}\centering factorized S-matrix\par [here, \dots]\end{minipage}&\begin{minipage}{7cm}\centering ? \par [e.g.\ $\superN=4$ SYM, ABJM, Fishnets]\end{minipage}\\\hline
\end{tabular}
\end{center}
\caption{Implications of local and nonlocal symmetries on the S-matrix.}
\label{tab:symresults}
\end{table}

In this paper we address this question for the case of $d=2$ spacetime dimensions, cf.\ \tabref{tab:symresults}. 
For explicitness, we restrict here to the case of nonlocal \emph{Yangian} charges, i.e.\ to integrable models of rational type, and investigate their impact on the $3\to 3$ particle S-matrix. It is well known that Yangian symmetry implies that the two-particle scattering matrix obeys the quantum Yang--Baxter equation (qYBE) \cite{Drinfeld:1985rx}. 
The three-particle S-matrix in a Yangian-invariant theory, however, could a priori split into a factorized part and an honest three-particle interaction  $\text R_{3\to3}$, cf.\ e.g.\ \cite{Iagolnitzer:1977sw}:
\begin{equation}
\label{eq:S33}
\text S_{3\to3}=\delta_{p_1,p_2,p_3}^{p_4,p_5,p_6} \prod_{jk} \text S^{jk}_{2\rightarrow 2}+\text R_{3\to 3}
=
\includegraphicsbox[scale=.7]{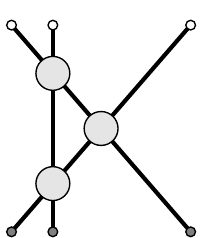}
+
\includegraphicsbox[scale=.7]{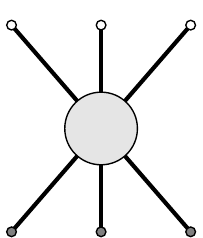}.
\end{equation}
In this notation, the question of whether Yangian symmetry implies factorized scattering turns into the question of whether $\text R_{3\to 3}=0$.

In the case of \emph{local} higher charges $\charge_k$ with $k>1$, an essential point in the argument leading to factorization is that these charges allow us to shift particle trajectories by a momentum (i.e.\ slope) dependent amount, such that multi-particle interactions can be decomposed into two-particle scattering processes, see e.g.\ \cite{Dorey:1996gd}:
\begin{align}
\includegraphicsbox[scale=.7]{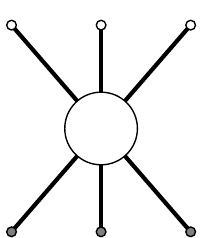}
\qquad \to\qquad
\includegraphicsbox[scale=.7]{FigYBELHS.pdf}
=
\includegraphicsbox[scale=.7]{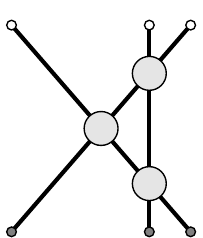}.
\end{align} 
Here the equality sign reflects the quantum Yang--Baxter equation that imposes consistency constraints on the two-particle scattering matrix.

Investigating the impact of nonlocal symmetries in the following, we will generically not assume that the sets of incoming and outgoing momenta or the corresponding rapidities are identical. However, in some cases the intricate constraint equations will not be tractable without this assumption.
One may wonder whether assuming conserved individual momenta allows to define a set of local conserved charges $\charge_k$ via their eigenvalue on one-particle states, schematically: 
\begin{equation}\label{eq:deflocs}
\charge_k\ket{p}\sim p^k \ket{p},
\end{equation}
and to proceed with the above arguments for factorized scattering following from local conserved charges. 
However, as pointed out in \cite{Dorey:1996gd}, it is not clear that given conserved sets of momenta, a local charge density exists which leads to an eigenvalue $\sum_{j=1}^n p_j^k$ on multi-particle states, generalizing \eqref{eq:deflocs}. Hence, conservation of the sets of momenta or rapidities does not yield a straightforward argument for the existence of local charges.

We will not investigate the consequences of nonlocal symmetries for (the absence of) particle production, but only study their impact on the $3\to3$ scattering matrix.
For a formal proof of the absence of particle production using nonlocal symmetries in the context of the nonlinear sigma model see L\"uscher's paper \cite{Luscher:1977uq} (cf.\ also \cite{Woo:1979nj}).%
\footnote{Note that there are examples of theories with a tower of nonlocal symmetries at the \emph{classical level} \cite{Curtright:1994be}, whose S-matrix features particle production \cite{Nappi:1979ig}.} 
While it might be reasonable to assume that in a theory with non-vanishing two-particle S-matrix, a factorized solution for the three-particle scattering exists, a priori it is not obvious that the factorized solution is compatible with the nonlocal symmetries. The paper \cite{Luscher:1977uq} also contains a discussion of this compatibility question for the nonlinear sigma model; a proof of factorized scattering based on nonlocal symmetries is not given there. 

In the following we will be interested in physical models with nonlocal symmetry algebras realized on asymptotic states. We assume that a representation $\gen{J}$ of the symmetry generators on asymptotic states exists, such that schematically
\begin{equation}
\comm{\gen{J}}{\Smat}=0.
\end{equation}
In particular, the above constraint equations will be specified to the case of the Yangian algebra and we will study the solutions $\Smat_{3\to 3}$ of these equations.
In doing so, we assume that the quantum space of asymptotic states is completely characterized by the Yangian, i.e.\ that all scattering states are labeled by the Lie algebra or Yangian quantum numbers.

While an interesting question on its own, one particular motivation for studying the implications of Yangian symmetry comes from the planar AdS/CFT correspondence. In this context integrable structures play a crucial role for the determination of involved observables and inspire the development of more and more refined tools \cite{Arutyunov:2009ga,Beisert:2010jr,Bombardelli:2016rwb}. The exact two-particle S-matrix of AdS/CFT can be written as a product of two two-particle S-matrices, each invariant under a centrally extended $\alg{su}(2|2)$ algebra in a so-called \emph{dynamic} representation \cite{Beisert:2005tm}. Moreover, the latter S-matrix is equivalent to Shastry's R-matrix \cite{Shastry:1986zz} for the Hubbard model \cite{Beisert:2006qh,Martins:2007hb}. 
While the importance of factorization of the higher-point S-matrix in AdS/CFT  has been discussed and checked in various contexts (see e.g.\ \cite{Staudacher:2004tk,Puletti:2007hq}), to our knowledge the relation of this fact to the underlying Yangian algebra has not been studied. It was shown, however, that the $\alg{su}(2|2)$ symmetric two-particle S-matrix is Yangian invariant \cite{Beisert:2007ds} (see also \cite{Arutyunov:2006yd}). 
This implies that it obeys the quantum Yang--Baxter equation and thus furnishes a building block that is consistent with factorization. 
Interestingly, the special dynamic representation of the Lie algebra symmetry $\alg{su}(2|2)$ allows to completely fix the \emph{two-particle} S-matrix up to an overall factor. For the bound-state S-matrix of \cite{Arutyunov:2008zt}, the full Yangian symmetry is required to fix all coefficients \cite{deLeeuw:2008dp}.

More recently, a bootstrap approach was suggested which is based on the idea to compose correlation functions using so-called hexagon form factors that obey certain axioms, see \cite{Komatsu:2017buu} for a recent pedagogical introduction. Also within this hexagon approach the exact two-particle S-matrix of AdS/CFT features prominently. In particular, it is conjectured that the hexagon amplitude essentially factorizes into the above $\alg{su}(2|2)$-invariant two-particle S-matrices \cite{Basso:2015zoa}. As suggested in that paper, it would be desirable to replace this assumption by the Yangian symmetry as an axiomatic input to the boostrap program.%
\footnote{We thank P.~Vieira for elaborating on this comment in their paper.}
This emphasizes the interest in understanding the connection between Yangian symmetry and factorized scattering, in particular for the case of the special dynamic representations important in the AdS/CFT correspondence.

In the following, we introduce the Yangian and its relation to the S-matrix and give a brief overview of our results in \secref{sec:YangianAndS} and \secref{sec:overview}, respectively. In \secref{sec:examples} we implement the Yangian constraints on the two- and three-particle S-matrices in \texttt{Mathematica} and study their solutions for the following three concrete examples:
\begin{enumerate}
\item \textbf{The fundamental representation of $\alg{su}(N)$:}\par
This representation of the Yangian plays a role
in various different contexts, e.g.\ for the Heisenberg spin chain or the chiral Gross--Neveu model.
Moreover, this case serves as a warm-up exercise for the more involved dynamical representations of AdS/CFT and to introduce our methodology. 
\item \textbf{The fundamental representation of $\alg{u}(1|1)$ (undynamic and dynamic):}\par
Adding supersymmetry to the above problem, we consider the dynamic representations of \cite{Beisert:2005wm} which are relevant to the scattering in the $\alg{su}(2|1)$ sector of $\superN=4$ SYM theory. While the dynamic property of a rapidity-dependent action of the symmetry generators is implemented, some of the complications of the full AdS/CFT scattering problem (e.g.\ braiding factors) are still projected out in this case. We also discuss the Yangian constraints in the undynamic limit which represents the scattering in conventional spin chain models.
\item \textbf{The fundamental representation of $\alg{su}(2|2)(\ltimes \mathbb{R}^2$) (undynamic and dynamic):}\par
Finally, we consider the full symmetry algebra underlying the scattering in AdS/CFT.  The new feature of its dynamic representation is given by non-trivial braiding factors that render the action of the Lie algebra generators (the level 0 of the Yangian) nonlocal. Again we consider the full dynamic representation, as well as its undynamic limit that plays a role in the condensed matter literature.

\end{enumerate}


\section{Nonlocal Yangian Symmetry and the S-matrix}
\label{sec:YangianAndS}

In this section we introduce the Yangian algebra underlying the subsequent analysis of symmetry constraints on the S-matrix. Moreover, we establish the role of the so-called \emph{dynamic} representations relevant in the context of the AdS/CFT duality and their relation to Hopf algebra twists of the Yangian.


\subsection{S-Matrices in Two Dimensions}

It is a special feature of theories in (1+1) dimensions that we may order particles with respect to their rapidities. We use this feature to define incoming and outgoing states in a scattering process. An incoming state of an $n$-particle scattering process is an asymptotic state denoted by
\begin{align}
&\ket{a_1,u_1;...;a_n,u_n}_\text{in},
&
&u_1>u_2>...>u_n,
\end{align}
where the particle of type $a_1$ moves behind the particle of type $a_2$ etc. The corresponding rapidities are ordered such that all particles participate in the scattering process. 
Similarly, our notion for an outgoing state of an $n$-particle scattering process is given by
\begin{align}\label{outgoing}
&\ket{a_1,u_1;...;a_n,u_n}_\text{out},
&
&u_1<u_2<...<u_n,
\end{align}
such that the particle labeled by $1$ moves behind particle $2$ etc., but the rapidities are now in the reverse order. Thus no scattering is possible for this configuration. 

Following the usual convention, the S-matrix is the operator that maps an $m$-particle outgoing state to an $n$-particle incoming state according to
\begin{align}
\ket{a_1,u_1;\dots;a_n,u_n}_\text{in}=\text S^{a_1 \dots a_n}_{b_1\dots b_m}(u_i,v_i)\ket{b_1,v_1;\dots;b_m,v_m}_\text{out}.\label{Sinout}
\end{align}
A priori, the S-matrix may depend on all the rapidities $u_i$ and $v_i$ of both states, and the sets of incoming and outgoing rapidities need not be equal.
We denote the single-particle Hilbert spaces of rapidity $u$ by $V_{u}$,%
\footnote{Below we will add further quantum numbers labeling these spaces.}
such that the S-matrix acts as
\begin{align}
\text S:V_{v_1}\otimes V_{v_2}\otimes ...\otimes V_{v_m} \rightarrow V_{u_1}\otimes V_{u_2}\otimes ...\otimes V_{u_n},
\label{eq:Smap}
\end{align}
with incoming rapidities $u_1>u_2>...>u_n$ and outgoing rapidities $v_1<v_2<...<v_m$.
Take as an example a free theory, i.e.\ the particles do not interact at all and they may only overtake each other. In this case, the S-matrix simply permutes the one-particle Hilbert spaces
and acts on a two-particle state as
$
\text S\ket{a_2,u_2;a_1,u_1}_\text{out}\sim\ket{a_1,u_1;a_2,u_2}_\text{in}.
$
We call an S-matrix that maps as \eqref{eq:Smap} with $n\equiv m$ an $n$-particle S-matrix.

\paragraph{Integrability.}
An integrable theory in two dimensions with a notion of scattering is characterized by the fact that the two-particle S-matrix obeys the quantum Yang--Baxter equation:
\begin{align}
\text S_{12}(u_1,u_2)\text S_{23}(u_1,u_3)\text S_{12}(u_2,u_3)=\text S_{23}(u_2,u_3)\text S_{12}(u_1,u_3)\text S_{23}(u_1,u_2).
\label{qYBE}
\end{align}
Here the indices on S denote the single-particle Hilbert spaces it non-trivially acts on.
Moreover,  there is no particle creation or annihilation and the individual momenta or rapidities are conserved. Finally, higher-point scattering matrices factorize into two-particle S-matrices, in particular the three-particle S-matrix obeys
\begin{align}
\text S_{123}(u_1,u_2,u_3)=\text S_{12}(u_1,u_2)\text S_{23}(u_1,u_3)\text S_{12}(u_2,u_3).\label{3fact}
\end{align}
Throughout this paper we will only investigate S-matrix elements that preserve the number of particles, i.e.\ $2\to 2$ and $3\to 3$ particle scattering. 
As discussed above, the vanishing of S-matrix elements with different numbers of incoming and outgoing particles is a generic feature of integrable models. Nevertheless, a formal proof of the absence of particle production based on the existence of quantum nonlocal charges is still lacking except for certain cases \cite{Luscher:1977uq}. 
\subsection{The Yangian Algebra}
\label{subsec:Yangian}

The infinite dimensional Yangian algebra $Y[\mathfrak g]$ is generated by two sets of generators $\levz^a$ and $\levo^a$.   The first set contains the so-called level-0 generators $\levz^a$ of a Lie algebra $\alg{g}$ with structure constants $f^{ab}{}_{c}$ and $a=1,...,\text{dim}(\mathfrak g)$. The second set of so-called level-1 generators~$\levo^a$ contains the same number of elements and obeys similar commutation relations as the Lie algebra generators:
\begin{align}
&[\levz^a,\levz^b]=f^{ab}{}_c\levz^c,
&
&[\levz^a,\levo^b]=f^{ab}{}_c\levo^c.
\label{eq:Yangianalg}
\end{align}
All higher levels are successively generated by commuting the generators of the previous levels. Thus, although the Yangian is an infinite-dimensional algebra, it is spanned by a finite set of operators --- the level-0 and level-1 generators.  This construction does not work for arbitrary Lie algebra representations, but depends on whether the additional \emph{Serre-relations} underlying the definition of the Yangian are satisfied:%
\footnote{See e.g.\ \cite{Drummond:2009fd} for the additional grading factors in the supersymmetric case. For $\alg{g}=\alg{sl}(2)$ a second Serre relation enters which is implied in the given relation for other algebras.}
\begin{align}
&[\levo^a,[\levo^b,\levz^c]]-[\levz^a,[\levo^b,\levo^c]]=\hbar^2 g^{abc}{}_{def}\{\levz^d,\levz^e,\levz^f\}.
\label{Serre}
\end{align}
Here we have abbreviated
\begin{align}
&g^{abc}{}_{def}=\tfrac 1{24}f^{ai}{}_{d}f^{bj}{}_ef^{ck}{}_ff_{ijk},
&
& \{X_1,X_2,X_3\}=\sum_{i\neq j\neq k}X_iX_jX_k.
\end{align}
The Yangian is defined as the algebra with $\hbar=1$ but it is illustrative to explicitly display the quantum parameter of this quantum group.


\paragraph{Coproduct Structure.}

The Yangian is a Hopf algebra with a coproduct structure. Given a representation of $\levz^a$ and $\levo^a$ acting on a one-particle state, the action on an $n$-particle state is determined by the coproduct $\Delta$ given by
\begin{align}
&\Delta^{n-1}(\levz^a)=\sum_{j=1}^n\levz_j^a,
&\Delta^{n-1}(\levo^a)=\sum_{j=1}^n\levo_j^a+\hbar \,f^{a}{}_{bc}\sum_{1\leq j<k\leq n}\levz_{j}^b \levz_{k}^c.
\label{coproduct}
\end{align}
Note that the level-1 coproduct in \eqref{coproduct} is sensitive to a rescaling of the generators, which allows to set $\hbar=1$.
The coproduct structure reveals that the Yangian is an algebra of nonlocal charges: While the coproduct of the level-0 generators acts locally on a single space (excitation), the level-$1$ generators act nonlocally on two spaces simultaneously. 


\paragraph{Evaluation Representation.}

In the context of scattering processes, the Yangian typically enters via its \emph{evaluation representation} $\rho_{u}$. The evaluation representation is obtained by lifting a representation $\rho$ of the underlying Lie algebra to a representation of the full Yangian algebra by setting
\begin{align}
&\rho_{u}(\levz^a)=\rho(\levz^a),
&
\rho_{u}(\levo^a)=u\rho(\levz^a)
\label{evrep}.
\end{align}
Here the evaluation parameter $u$ is proportional to the rapidity variable of the model and we do not distinguish between those quantities in the following.
The role of the rapidity can be understood by noting that the Yangian has an external automorphism $B_u$ which is usually realized by the Lorentz boost in two-dimensional Yangian symmetric field theories, cf.\ e.g.\ \cite{Loebbert:2016cdm}:
\begin{align}
\label{eq:boost}
B_u(\levz_a)&=\levz_a,
&
B_u(\levo_a)&=\levo_a+u\levz.
\end{align}
In order to furnish a representation of the Yangian algebra, the representation $\rho_{u}$ has to satisfy the Serre-relations \eqref{Serre} which constrains the underlying Lie algebra representation $\rho$. 
The action of the Yangian evaluation representation on multi-particle states follows from the coproduct \eqref{coproduct} according to
\begin{align}
&\rho_{u}^{\otimes n}(\Delta^{n-1}\levz^a)\ket{a_1,u_1;...;a_n,u_n}_{\text{in}\above0pt\text{out}}=\sum_{j=1}^n\rho^{\otimes n}(\levz^a_j)\ket{a_1,u_1;...;a_n,u_n}_{\text{in}\above0pt\text{out}}\nonumber\\ 
&\rho_{u}^{\otimes n}(\Delta^{n-1}\levo^a)\ket{a_1,u_1;...;a_n,u_n}_{\text{in}\above0pt\text{out}}\nonumber\\
&\hspace{1cm}=\left(\sum_{j=1}^n u_j\rho^{\otimes n}(\levz ^a_j)+
\hbar\,
 f^{a}{}_{bc}\sum_{1\leq j<k\leq n}\rho^{\otimes n}(\levz^b_{j})\rho^{\otimes n}(\levz^c_{k})\right)\ket{a_1,u_1;...;a_n,u_n}_{\text{in}\above0pt\text{out}}.\label{msiteevrep}
\end{align}
Here $\rho_{u}^{\otimes n}$ and $\rho^{\otimes n}$ denote the representations on a tensor product of length $n$. In the following we will generically not distinguish between the symbols for the abstract algebra element $\levz$ and its representation $\rho(\levz)$.

\subsection{Dynamic Representations and Braiding}
\label{subsecDynUndyn}

A Lie algebra symmetry is typically realized on a one-particle excitation with rapidity $u$ according to
\begin{equation}
\gen{J}^a \ket{a_1,u} = j^a \ket{a_2,u},
\end{equation}
where $a$ enumerates the generators.
Here $a_1,a_2$ denote the type of excitations and $j^a$ represents some generator-dependent function.
Conventionally $j^a$ is constant, i.e.\ independent of the rapidity $u$, and $\gen{J}^a$ has a trivial (local) tensor product action on multi-particle states:
\begin{equation}
\gen{J}^a \ket{a_1,u_1;\dots;a_n,u_n}=\sum_{k=1}^n\gen{J}_{k}^a \ket{a_1,u_1;\dots;a_n,u_n}.
\end{equation}
 We will refer to this type of symmetry representation as \emph{undynamic}. 

In the context of the AdS/CFT correspondence, so-called \emph{dynamic} representations play an important role. These are representations with $j^a=j^a(u)$ and typically a nonlocal coproduct for level-0 generators with a so-called braiding factor $\mathcal{U}$: 
\begin{equation}\label{eq:twistedcoproduct0}
\Delta \gen{J}_a=\gen{J}_a\otimes \idop +\,\braiding^{[a]} \otimes \gen{J}_a.
\end{equation}
The presence of a non-trivial braiding implies that even the level-0 symmetry acts nonlocally.
Also the Yangian level-1 generators inherit a braiding according to (see e.g.\ \cite{Beisert:2007ds}):
 \begin{equation}\label{eq:twistedcoproduct1}
 \Delta \levo_a=\levo_a \otimes \idop + \braiding^{[a]}\otimes \levo_a+\hbar\, f_a{}^{bc} \levz_b\, \braiding^{[c]}\otimes \levz_c.
 \end{equation}

 
\paragraph{Length Changing and Braiding.}
Dynamic representations can be alternatively considered as either length-changing Lie algebra representations or as Hopf algebra representations containing an additional generator (the above braiding factor), which renders the coproduct of Lie algebra generators nonlocal and the single-particle action rapidity dependent \cite{Beisert:2005tm,Gomez:2006va,Plefka:2006ze}.
Consider for instance some excitation $\Phi$ with momentum $p$ over an infinite spin chain vacuum composed of scalar fields $Z$:
\begin{equation}\label{eq:onepartpsi}
\ket{\Phi_p}=\sum_k e^{ipk} \ket{\dots ZZ\markarrow{\Phi}{k} ZZ\dots}=
 \includegraphicsbox{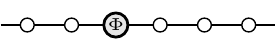}.
\end{equation}
Let the action of the supersymmetry generator $\gen{\tilde Q}$ on a fermionic excitation (i.e.\ $\Phi=\psi$) be given by
\begin{equation}\label{eq:Qtilde}
 \gen{\tilde Q} \ket{\psi_p}=\ket{Z\phi_p }=
 \includegraphicsbox{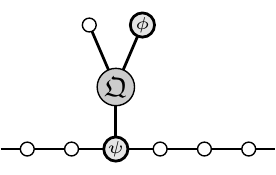},
\end{equation}
where $\phi$ denotes a bosonic field.
The generator $\gen{\tilde Q}$ is length changing since it inserts a scalar vacuum field (a marker) $Z$ in front of the excitation. If we look at asymptotic states defined on an infinite spin chain, the marker has no effect on one-particle states of the form \eqref{eq:onepartpsi} with $\ket{Z\phi_p }\simeq \ket{\phi_p }$.
This is different once we consider the action on multi-particle states, e.g.\
\begin{equation}\label{eq:lengthtodynamic}
\Delta\gen{\tilde Q} \ket{\psi_{p_1}\psi_{p_2}}= \ket{Z\phi_{p_1} \psi_{p_2}}+ \ket{\psi_{p_1} Z\phi_{p_2}}.
\end{equation}
Moving the marker past the excitation to the very left corresponds to an index shift and thus to a relative phase $e^{ip_1}$ between the two summands in \eqref{eq:lengthtodynamic}. We may thus get rid of the notion of length changing and instead define a non-trivial braided coproduct 
\begin{align}
&\Delta \gen{Q}=\gen{Q}\otimes \idop +\,\braiding^{[\gen{Q}]} \otimes \gen{Q},
&
&\braiding^{[\gen{Q}]}=e^{ip},
\end{align}
such that \cite{Gomez:2006va,Plefka:2006ze}
\begin{equation}
\Delta\gen{ Q} \ket{\psi_{p_1}\psi_{p_2}}= \ket{\phi_{p_1} \psi_{p_2}}+ e^{ip_1} \ket{\psi_{p_1} \phi_{p_2}}.
\end{equation}
Here the untilded operator $\gen{Q}$ is defined without the length-changing effect as opposed to \eqref{eq:Qtilde}:
\begin{equation}
\gen{Q}\ket{\psi_p}=\ket{\phi_p}.
\end{equation}
The new generator $\braiding^{[\gen{Q}]}$ implies that Lie algebra generators acquire a nonlocal coproduct. In the case of the fundamental representation of $\alg{su}(2|2)$ considered in \secref{sec:su22}, introducing this generator furthermore requires to centrally extend the algebra to $\alg{su}(2|2)\ltimes \mathbb{R}^2$.%
\footnote{See \cite{Beisert:2014hya} for an interpretation of the braiding factor in terms of the RTT-realization of the Yangian.}

\paragraph{Hopf Algebras and Scattering Matrices.}

A Hopf algebra $\mathfrak{a}$ with coproduct $\Delta$ and with a universal R-matrix $\uR$ is called quasi-triangular if we have
\begin{align}\label{eq:deltaR}
(\Delta\otimes \idop)(\uR)&=\uR_{13}\uR_{23},
&
(\idop\otimes\Delta)(\uR)=\uR_{13}\uR_{12},
\end{align}
as well as co-commutativity:
\begin{align}\label{eq:cocomm}
\Delta^\text{op}(X)\uR&=\uR \Delta(X),
&
&X\in \alg{a}.
\end{align}
Here the opposite coproduct $\Delta^\text{op}=\permop\Delta \permop$ is defined as the coproduct $\Delta$ with flipped tensor factors.
The above relations imply that $\uR$ obeys the quantum Yang--Baxter equation. The universal R-matrix $\uR$ represents a formal expression for the above intertwiner in terms of the (infinitely many) algebra generators that upon choosing a representation $\rho$ turns into the two-particle S-matrix via 
\begin{equation}\label{eq:SfromR}
\Smat=\permop \rho\otimes\rho(\uR),
\end{equation} 
with $\permop$ the permutation operator swapping the tensor factors. The Yangian is not quasi-triangular. 
Still the Yangian obeys relations similar to \eqref{eq:cocomm}:
\begin{align}\label{eq:boostR}
(B_u\otimes \idop)\Delta^\text{op}(X) &=\uR(u)(B_u\otimes \idop)\Delta(X) \uR^{-1}(u),
&
X\in Y[\alg{g}].
\end{align}
These imply that on the level of representations, the Yangian has quasi-triangular properties. Here $B_u$ denotes the boost automorphism \eqref{eq:boost}.

According to a theorem by Drinfeld \cite{Drinfeld:1985rx} there is a unique rational series $\mathcal{R}(u)=\sum_k u^{-k} \mathcal{R}_k$ with expansion coefficients $\mathcal{R}_k\in Y[\alg{g}]\otimes Y[\alg{g}]$ that obeys the relations \eqref{eq:deltaR, eq:boostR} and whose expansion is given in terms of the Yangian generators as:
\begin{equation}
\label{eq:expR}
\log \mathcal{R}(u)=\sfrac{1}{u}\levz_a\otimes\levz^a+\sfrac{1}{u^2}\big(\levo_a\otimes\levz^a-\levz_a\otimes \levo^a\big)+\mathcal{O}\big(\sfrac{1}{u^3}\big).
\end{equation}
This \emph{pseudo-universal} R-matrix obeys the quantum Yang--Baxter equation. Hence, choosing a representation $\rho$ yields via \eqref{eq:SfromR} an associated two-particle S-matrix that induces a consistent factorized solution for three-particle scattering. Note that in principle a prescription exists to construct the pseudo-universal R-matrix using the so-called Yangian double, cf.\ \cite{Khoroshkin:1994uk,2005math......4302S,Rej:2010mu}. 
A universal R-matrix which reproduces the AdS/CFT S-matrix when evaluated on the fundamental representation, however, is not known, see e.g.\ \cite{Moriyama:2007jt,Beisert:2016qei} for related comments.

\paragraph{Braiding and Twist.}
The introduction of braiding factors $\mathcal{U}$ can formally be considered as a Reshetikhin twist defined in \cite{Reshetikhin:1990ep}. This twisting implies that even the level-zero symmetry generators are nonlocal.
A Reshetikhin twist is a modification of a quasitriangular Hopf algebra via some twist operator $F$, which obeys the quantum Yang--Baxter equation 
\begin{equation}
F_{12}F_{13}F_{23}= F_{23}F_{13}F_{12},
\end{equation}
as well as a unitarity relation
$
F_{12}F_{21}=1,
$
and the axioms
\begin{align}
(\Delta\otimes \idop)(F)&=F_{13}F_{23},
&
(\idop\otimes\Delta)(F)=F_{13}F_{12}.
\end{align}
Then the twisted Hopf algebra is also quasitriangular with coproduct and universal R-matrix given by
\begin{align}
\Delta^{(F)}(X)&=F\Delta(X)F^{-1},
&
\uR^{(F)}_{12}=F_{21}\uR_{12} F^{-1}_{12}.
\end{align}
As noted in \cite{Spill:2012qe}, the above braiding factors $\braiding \simeq e^{ip}$ can be obtained by a formal twist generating the dynamic representation of $\alg{su}(2|2)$ (the symmetry of the AdS/CFT S-matrix):
\begin{equation}
F=\exp\big(i\sfrac{p}{2} \otimes \gen{A}\big).
\end{equation}
Here we define the coproduct of the formal momentum generator as $\Delta p=p\otimes \idop +\idop \otimes p$ and $\gen{A}$ is one of three outer automorphisms of the triple centrally extended algebra $\alg{psu}(2|2)$ that together generate an $\alg{sl}(2)$ algebra, i.e.\ one considers $\alg{sl}(2)\ltimes\alg{psu}(2|2)\ltimes \mathbb{R}^3$.
\section{Constraint Relations and Summary of Results}
\label{sec:overview}

Yangian symmetry has important implications on the S-matrix which relates multi-particle states in the in- and out-state basis. The conservation of the Yangian charges in a scattering process constrains the $n$-particle S-matrix according to
\begin{align}
&\text{level 0:}\ \ [\Delta^{n-1}\levz^a,\text S]=0\label{level0},\\
&\text{level 1:}\ \ [\Delta^{n-1}\levo^a,\text S]=0.\label{level1}
\end{align}
The coproduct entering the constraint equations \eqref{level0,level1} is specified in 
\eqref{coproduct}.
Higher levels of Yangian generators do not impose further constraints on the S-matrix.

We will investigate these constraints and their implications on S-matrices for the fundamental representations $\rho$ of the three different Yangian algebras $Y[\mathfrak{su}$(N)], $Y[\mathfrak{u}$(1|1)] and $Y[\mathfrak{su}(2|2)\ltimes \mathbb R^2]$ in detail in the following sections. Doing so we focus on two and three-particle S-matrices. In particular, we wish to clarify the question of whether the Yangian symmetry implies that the three-particle S-matrix necessarily factorizes into two-particle S-matrices. 
Explicit expressions of the coproducts for the different algebras can be found in \secref{sec:examples}.


\subsection{Two-Particle S-Matrix and Difference Form}
\label{sec:diffform}

The above Yangian constraints imply that an invariant two-particle S-matrix obeys the qYBE \eqref{qYBE}.
Moreover, for the case of an (untwisted) Yangian algebra, the symmetry-invariant two-particle S-matrix typically only depends on the difference of rapidity parameters.%
\footnote{In a Lorentz-invariant theory with one spatial direction, this statement reflects the boost invariance of the S-matrix provided that the rapidities are additive under Lorentz boosts.}
In fact, this feature directly follows from the Yangian constraint equations for the S-matrix \eqref{level0,level1}, cf.\ e.g.\  \cite{Chari:1994pz,Beisert:2007ds}. The level-1 constraint \eqref{level0} for an arbitrary level-0 generator $\levz$ and a corresponding level-1 generator $\levo$ is given by
\begin{align}
\comm{\Delta\levo}{\text S_{12}}=\left(u_1 \levz_1+u_2 \levz_2\right)\text S_{12}-\text S_{12}\left(v_1\levz_1+v_2 \levz_2 \right)+\left[\levz\otimes\levz,\text S_{12}\right]=0.
\end{align}
Here $\levz\otimes\levz$ shall represent the bilocal term of the coproduct \eqref{coproduct}. This constraint can be rearranged using the level-0 constraint \eqref{level0}, i.e.\
$
\left[\levz_1+\levz_2,\text S_{12}\right]=0
$,
into the form
\begin{align}\label{rapdifuv}
u_{12}\levz_1\text S_{12}
-(v_1-u_2)\text S_{12}\levz_1
-(v_2-u_2)\text S_{12}\levz_2
+\left[\levz\otimes\levz,\text S_{12}\right]=0,
\end{align}
where $u_{ij}:=u_i-u_j$. Hence, for unequal sets of incoming and outgoing rapidities $\{u_i\}\neq\{v_i\}$, the two-particle S-matrix depends only on three independent rapidity parameters.

For the case of equal sets of incoming and outgoing rapidities, i.e.\ for $\{u_i\}=\{v_i\}$, the above relation \eqref{rapdifuv} simplifies to
\begin{align}\label{rapdif}
u_{12}\left(\levz_1\text S_{12}-\text S_{12}\levz_2\right)+\left[\levz\otimes\levz,\text S_{12}\right]=0.
\end{align}
Thus the rapidity-dependence of the two-particle S-matrix is governed by an equation that contains only the rapidity difference, which implies (up to an overall factor) the difference form of the S-matrix:
\begin{equation}
\text S_{12}(u_1,u_2)=\text S_{12}(u_{12}).
\label{eq:diffform}
\end{equation} 
This observation can be attributed to the external boost automorphism of the Yangian as defined in \eqref{eq:boost}, i.e.\ Lorentz boost symmetry is tightly connected to conventional Yangian invariance. Note however that the above arguments do not apply to the braided coproducts \eqref{eq:twistedcoproduct0,eq:twistedcoproduct1} which underlie the scattering in AdS/CFT. Hence, the S-matrix of AdS/CFT is not of difference form \cite{Beisert:2005tm}. We give explicit examples in \secref{sec:examples}.


\subsection{Three-Particle S-Matrix and Factorized Scattering}
\label{subsec:Table}

In general, a three-particle S-matrix is of the form \eqref{eq:S33}. As discussed above, we check here whether Yangian symmetry implies the absence of an honest three-particle interaction $R_{3\rightarrow 3}$. This is equivalent to the three-particle S-matrix factorizing as 
in \eqref{3fact}
and the two-particle S-matrix satisfying the qYBE \eqref{qYBE}.
Our considerations are grounded on the explicit implementation of the Yangian constraints \eqref{level0,level1} for the three-particle S-matrix with all scattered particles transforming in one of the symmetry representations specified in the introductory \secref{sec:intro}. In the following we summarize the results of our explicit implementation of Yangian constraints and their analysis in \texttt{Mathematica}. We give a more detailed discussion of this analysis in the following \secref{sec:examples}.


\paragraph{Results of Computer Implementation.}
We display the key results of our implementation in \tabref{table:results}. 
First of all we examined the Yangian constraints on S-matrices to check whether they imply that the sets of rapidities are conserved in two- and three-particle scattering processes, cf.\ the columns labeled by $\{u_i\}\overset{?}{=}\{v_i\}$ in \tabref{table:results}. As expected, in all cases conservation of the set of rapidities solves the constraint equations (with further constraints on the S-matrix discussed in the following). In those cases that are not marked by a star, we were able to show without further complications that this solution is also the unique solution to the Yangian constraints. Note that in addition to the level-0 constraints, we also require the level-1 constraints to deduce that the set of rapidities is conserved. In the cases marked by stars our implementation could not solve the involved constraint equations for generic algebra parameters/functions and different sets of incoming and outgoing rapidities. In the cases denoted by a box we were not able to solve the constraint equations at all. In the other cases, these complications could be circumvented by

\begin{table}
\renewcommand{\arraystretch}{1.1}
\begin{tabular}{|cc||c|c|c|c|c|}\hline
 & & $Y[\mathfrak{su}(N)]$ & \multicolumn{2}{c|}{$Y[\mathfrak{u}(1|1)]$} & \multicolumn{2}{c|}{$Y[\mathfrak{su}(2|2)]$} \\
 & & \hphantom{undynamic} & undynamic & \hphantom{u}dynamic\hphantom{n} & undynamic & dynamic ($\ltimes\mathbb R^2$)\\
\hline\hline
\parbox[t]{2mm}{\multirow{4}{*}{\rotatebox[origin=c]{90}{$2\rightarrow 2$}}} & $\{u_i\}\overset{?}{=}\{v_i\}$&\checkmark & \checkmark & (\checkmark)$^{\star}$ & \checkmark & (\checkmark)$^\star$\\
 & S from level 0 &\checkmark [2dof] & \checkmark [2dof]&\checkmark [2dof] & \checkmark [2dof] & \checkmark [1dof] \\
 & S from level 1 &\checkmark [1dof] & \checkmark [1dof] &\checkmark [1dof] & \checkmark [1dof] & \checkmark [1dof] \\
 & qYBE &\checkmark & \checkmark & \checkmark & \checkmark & (\checkmark)$^{\star\star\star}$ \\
 \hline
\parbox[t]{2mm}{\multirow{4}{*}{\rotatebox[origin=c]{90}{$3\rightarrow 3$}}}  & $\{u_i\}\overset{?}{=}\{v_i\}$&\checkmark & (\checkmark)$^{\star\star}$ & $\Box$  & \checkmark &  $\Box$\\
 & S from level 0 &\checkmark [5/6dof] & \checkmark [6dof] & \checkmark [6dof] & \checkmark [10dof] & (\checkmark [2dof])$^{\star\star\star}$ \\
 & S from level 1 &\checkmark [1dof] & \checkmark [1dof] & \checkmark [1dof] & \checkmark [1dof] & (\checkmark [1dof])$^{\star\star\star}$ \\
 & factorization &\checkmark & \checkmark & \checkmark & \checkmark & (\checkmark)$^{\star\star\star}$\\\hline
\end{tabular}
\caption{Results of the analysis of Yangian constraints corresponding to the fundamental representation of the Lie algebras $\mathfrak{su}(N)$, $\mathfrak{u}(1|1)$ and $\mathfrak{su}(2|2)\ltimes\mathbb R^2$}
\label{table:results}
\end{table}
\begin{itemize}
\item[$\star$] further specifications: 
Here we exploited the conservation of the total momentum and the energy in the form of the concrete AdS/CFT model. The latter corresponds to the conservation of the rapidity dependent eigenvalue of a dynamic level-0 Yangian generator. We solved these equations for random real numerical values of the incoming momenta and the coupling $g$, which enters the energy formula.
\item[$\star\star$] restricting to exemplary sets of numerical values for the level-0 eigenvalues: Here we facilitated the solution of the constraint equations by setting the (rapidity independent) eigenvalues of the level-0 generators to numerical values. 
\end{itemize}

In addition to constraints on the rapidities, the Yangian imposes constraints on the form of the S-matrix. For the analysis of the latter, we now assumed that the set of rapidities is conserved. In all considered cases, the Yangian constraints then fix the two- and three-particle S-matrices uniquely up to an overall factor, where in the cases marked by
\begin{itemize}
\item[$\star\star\star$] we solved the constraint equations using generic sets of numerical values for the rapidities and coupling constant.
\end{itemize}
Note that in  \tabref{table:results} we specify the number of degrees of freedom (dof) of the S-matrix after applying the level-0 (labeled by ``S from level 0'') and both level-0 and level-1 Yangian constraints (labeled by ``S from level 1'').\footnote{In the three-particle scattering case for $Y[\mathfrak{su}(N)]$ the number of degrees of freedom after exploiting the level-0 constraint depends on the rank $N$ of the Lie algebra. For $N=2$ there are 5 independent coefficients in the $\mathfrak{su}(2)$-invariant S-matrix, whereas this number is 6 for $N>2$.} In all considered cases the S-matrix was uniquely determined by the Yangian constraints up to a single degree of freedom, i.e.\ the overall factor of the S-matrix. 
Finally, we checked whether the resulting S-matrices are consistent with factorized scattering which we did by verifying \eqref{qYBE,3fact}. These equations were satisfied in all discussed cases. Again in all cases marked by $\star\star\star$, we did this analysis by using exemplary numerical values for the rapidities and coupling constant. We specify the references to related results in the literature in the explicit discussion of our analysis in \secref{sec:examples}.


\section{Explicit Implementation of Yangian Constraints}
\label{sec:examples}

In this section we outline the concrete analysis of the Yangian constraints \eqref{level0} and \eqref{level1} on S-matrices for $Y[\mathfrak{su}(N)]$, $Y[\mathfrak{u}(1|1)]$ and $Y[\mathfrak{su}(2|2)\ltimes\mathbb R^2]$ whose results we already presented in Table \ref{table:results}. Doing so, we elaborate on the algebras and the representations we considered, their coproduct structure and further ingredients that were necessary in this analysis.

Let us briefly motivate the choice of considered algebras. A good starting point for the concrete evaluation of the constraint equations \eqref{level0} and \eqref{level1} is the Yangian $Y[\mathfrak{su}(N)]$. At level 0 it only contains a single type of generator and its fundamental representation can be implemented straight-forwardly into \texttt{Mathematica} which we used to analyze the constraint equations. We outline the main steps of the analysis in \secref{subsec:suN}. Then in \secref{subsec:u11} we move on to the Yangian corresponding to the Lie superalgebra $\mathfrak{u}(1|1)$. It contains both commuting and anti-commuting generators, but only a single boson and a single fermion in its fundamental representation. Thus it is a good intermediate step before the investigation of $\mathfrak{su}(2|2)$ and its central extension. Finally, in \secref{sec:su22} we look at the Yangian constraints of the latter algebra which is particularly relevant in the context of the AdS/CFT correspondence because it appears as the symmetry algebra of the S-matrix of the $\mathfrak{psu}(2,2|4)$ spin chain of $\mathcal N=4$ SYM theory.

Note that we discuss the Yangian constraints to both Lie superalgebras in an undynamic and a dynamic setup as introduced in \secref{subsecDynUndyn}. While the S-matrices obtained from the dynamic representations are interesting in the context of the AdS/CFT correspondence, the undynamic cases are relevant for conventional condensed matter spin chains.

Furthermore note that for all Yangian algebras and their representations studied here we assume that the Serre relations \eqref{Serre} are satisfied and concentrate our studies on the analysis of the Yangian constraints \eqref{level0} and \eqref{level1}. For simplicity we remove the $\hbar$-dependence appearing in front of the nonlocal part in the Yangian coproduct by rescaling the generators.

We restrict the analysis of the Yangian constraints to the two- and three-particle scattering case with no particle creation or annihilation. Thus we do not check here whether the Yangian algebra implies the absence of particle creation or annihilation; L\"uscher showed that the nonlocal charges of the non-linear sigma model imply the absence of particle production \cite{Luscher:1977uq}.  
Another important feature of integrable theories in (1+1) dimensions is that the sets of incoming and outgoing rapidities in scattering processes are equal. Nevertheless, a priori we do not postulate this feature here but check whether Yangian symmetry yields it automatically;  however, not in all cases the resulting constraint equations could be solved, cf.\ \tabref{table:results}. Thus we look at the elements of the S-matrix that map the tensor product of $n$ one-particle Hilbert spaces $V_u$ labeled by the rapidity $u$ of the particle via 
\begin{align}
\text S(u_i;v_i): V_{v_1}\otimes V_{v_2}\otimes\dots\otimes V_{v_n}\rightarrow V_{u_1}\otimes V_{u_2}\otimes\dots\otimes V_{u_n}. 
\label{asym}
\end{align}
Here $\{v_1, v_2,...,v_n\}$ denotes the set of rapidities of the outgoing and $\{u_1,u_2,...,u_n\}$ the set of rapidities of the incoming particles in a scattering event. The rapidities shall be ordered as $v_1<v_2<...<v_n$ and $u_1>u_2>...>u_n$ as before.

\subsection{Yangian Constraints for $\alg{su}(N)$}
\label{subsec:suN}

As a first exercise we consider the Yangian constraints in the evaluation representation for the fundamental representation of $\alg{su}(N)$.
Note that the two-particle S-matrix for the fundamental representation of $\alg{su}(N)$ can also be obtained from the known universal R-matrix \cite{Rej:2010mu} and satisfies the qYBE \eqref{qYBE} automatically, cf.\ the discussion in \secref{subsecDynUndyn}.


\paragraph{$Y[\mathfrak{su}(N)]$ in the Evaluation Representation.}

The level-0 generators of the Yangian $Y[\mathfrak{su}(N)]$ are the generators $\mathfrak R^a_b$ of the Lie algebra $\mathfrak{su}(N)$. They satisfy the commutation relations
\begin{align}
&[\mathfrak R^a_b,\mathfrak R^c_d]=\delta_b^c\mathfrak R^a_d-\delta^a_d\mathfrak R^c_b,
&
& a,b=1,\dots,N,
\label{LieComm}
\end{align}
and are traceless. The level-1 generators $\widehat{\mathfrak R}^a_b$ satisfy similar commutation relations via \eqref{eq:Yangianalg}. In the evaluation representation they act as defined in \eqref{evrep} which requires to fix the Lie algebra representation. As mentioned before we choose the fundamental representation of $\mathfrak{su}(N)$ which consists of $N$ bosons $\phi^a, a=1,...,N$. The evaluation representation requires us to further label each state by its evaluation parameter $u$ and thus we denote a one-particle state by $\ket{\phi^a,u}$. These states shall form an orthonormal set, i.e.\ $\braket{\phi^a,u|\phi^b,v}=\delta^b_a\delta_{u,v}$.\footnote{Note that we use the Kronecker delta $\delta_{u,v}$ regardless of whether the set of allowed rapidities is continuous or discrete. In case the rapidities are not quantized, we identify the symbol with the Dirac delta function, i.e.\ $\delta_{u,v}\equiv\delta(u-v)$.} Then the Yangian level-0 and -1 generators act in Dirac notation as
\begin{align}
&\mathfrak R^a_b\ket{\phi^c,u}=\delta^c_b\ket{\phi^a,u}-\tfrac 1N\delta^a_b\ket{\phi^c,u},
&
&\widehat{\mathfrak R}^a_b\ket{\phi^c,u}=iu \mathfrak R^a_b\ket{\phi^c,u}.
\label{sugenerators}
\end{align} 


\paragraph{\texorpdfstring{$Y[\mathfrak{su}(N)$}{[Y[su](N)}]-Constraints on the S-Matrix.}

In order to make the constraint equations \eqref{level0} and \eqref{level1} on $\Smat$ explicit, one needs the coproduct structures of the Yangian generators. They are given by
\begin{align}
\Delta\mathfrak R^a_b&=\mathfrak R^a_b\otimes 1+1\otimes\mathfrak R^a_b,\nonumber\\
\Delta\widehat{\mathfrak R}^a_b&=\widehat{\mathfrak R}^a_b\otimes 1+1\otimes\widehat{\mathfrak R}^a_b+\tfrac 12\mathfrak R^a_c\otimes\mathfrak R^c_b-\tfrac 12\mathfrak R^c_b\otimes\mathfrak R^a_c.
\end{align}
These equations generalize to higher order coproducts via \eqref{coproduct}. 
The explicit constraint equations for the $Y[\mathfrak{su}(N)$]-invariant $n$-particle S-matrix S$_{12...n}$ then take the form
\begin{align}
&\text{level 0:}\hspace{0.2cm}\left[\sum_{j=1}^n (\mathfrak R^a_b)_j,\text S_{12...n}(u_i,v_i)\right]=0,\label{Level0sun}\\
&\text{level 1:}\hspace{0.2cm}\left(i\sum_{j=1}^n u_n(\mathfrak R^a_b)_j+\frac 12\sum_{1\leq j<k\leq n}\left((\mathfrak R^a_c)_{j}(\mathfrak R^c_b)_{k}-(\mathfrak R^c_b)_{j}(\mathfrak R^a_c)_{k}\right)\right)\text S_{12...n}(u_i,v_i)=\nonumber\\
&\hspace{2.5cm}\text S_{12...n}(u_i,v_i)\left(i\sum_{j=1}^n v_j(\mathfrak R^a_b)_j+\frac 12\sum_{1\leq j<k\leq n}\left((\mathfrak R^a_c)_{j}(\mathfrak R^c_b)_{k}-(\mathfrak R^c_b)_{j}(\mathfrak R^a_c)_{k}\right)\right)\label{Level1sun}
\end{align}
in the evaluation representation \eqref{sugenerators}.
Again the $u_i$ represent the rapidities of the incoming and the $v_i$ those of the outgoing particles. 
In the following we discuss the results of our implementation of the above constraints for $n=2,3$ in \texttt{Mathematica}. 


\paragraph{Two-Particle S-Matrix.}

The (level-0) Lie algebra symmetry, here given by the fundamental representation of $\mathfrak{su}(N)$, constrains the invariants to all possible permutations acting on the tensor factors. 
Let us introduce a diagrammatic way to represent the action of the two-particle invariant $\mathcal I_{12}$ using permutation diagrams of the form
\begin{equation}
\mathcal I_{12}(u_{1,2};v_{1,2})=A_{12}
\begin{tikzpicture}[thick,scale=0.7, every node/.style={scale=0.7},baseline={([yshift=-.5ex]current bounding box.center)}]
\node at (2,0) [circle,fill,inner sep=1.2pt,label=below:$v_1$] {};
\node at (2.5,0) [circle,fill,inner sep=1.2pt,label=below:$v_2$] {};
\node at (2,1) [circle,fill,inner sep=1.2pt,label=above:$u_1$] {};
\node at (2.5,1) [circle,fill,inner sep=1.2pt,label=above:$u_2$] {};
\draw (2,0) to[out=95, in=275] (2.5,1);
\draw (2.5,0) to[out=85, in=265] (2,1);
\end{tikzpicture}
+B_{12}
\begin{tikzpicture}[thick,scale=0.7, every node/.style={scale=0.7},baseline={([yshift=-.5ex]current bounding box.center)}]
\node at (4,0) [circle,fill,inner sep=1.2pt,label=below:$v_1$] {};
\node at (4.5,0) [circle,fill,inner sep=1.2pt,label=below:$v_2$] {};
\node at (4,1) [circle,fill,inner sep=1.2pt,label=above:$u_1$] {};
\node at (4.5,1) [circle,fill,inner sep=1.2pt,label=above:$u_2$] {};
\draw (4,0) to[out=90, in=270] (4,1);
\draw (4.5,0) to[out=90, in=270] (4.5,1);
\node at (4.7,0.4) {$.$};
\end{tikzpicture}\label{sunl02}
\end{equation}
Here the two dots at the bottom of each diagram represent the particles before the action of the operator, i.e.\ in case of the S-matrix corresponding to the outgoing state with rapidities $v_1<v_2$. The state after the action of the operator is represented by the top dots which are associated to the incoming state with rapidities $u_1>u_2$ for the S-matrix. The level-0 constraints do not restrict the momenta of the particles which is why at this point the sets of rapidities $\{u_1,u_2\}$ and $\{v_1,v_2\}$ are allowed to be different.\footnote{In principle these rapidities are constrained by momentum conservation. Nevertheless, we do not impose this conservation law here since the relation $p=p(u)$ between the momentum $p$ and rapidity $u$ is model-dependent. In this analysis we only impose Yangian symmetry.} The lines represent the action of the corresponding permutation operators and connect particles of the same type. Thus, the first diagram represents a scattering event where the faster particle passes through the slower one, while the second diagram corresponds to a reflection. Accordingly, $A_{12}$ is the transmission coefficient and $B_{12}$ the reflection coefficient of this scattering event. 

Evaluating the level-1 constraints \eqref{Level1sun} explicitly yields two solutions for the coefficients and rapidities of the ansatz \eqref{sunl02}. 
The first one fixes the coefficients and the outgoing rapidities as
\begin{align}
&v_1=u_1,
&
&v_2=u_2,
& A_{12}=0,
& 
& B_{12}=\mathcal I_{12}^0\delta_{v_1,u_1}\delta_{v_2,u_2}.
\end{align}
Here $\mathcal I_{12}^0$ is an undetermined function depending on the free rapidities.
This solution is not compatible with the condition $u_1>u_2$ and $v_1<v_2$ for scattering events. Indeed, it does not account for the scattering of particles, but rather corresponds to an operator proportional to the identity:
\begin{equation}
\mathcal I_{12}^{(*)}(u_{1,2};v_{1,2})=\mathcal I_{12}^0\delta_{v_1,u_1}\delta_{v_2,u_2}
\begin{tikzpicture}[thick,scale=0.7, every node/.style={scale=0.7},baseline={([yshift=-.5ex]current bounding box.center)}]
\node at (4,0) [circle,fill,inner sep=1.2pt,label=below:$u_1$] {};
\node at (4.5,0) [circle,fill,inner sep=1.2pt,label=below:$u_2$] {};
\node at (4,1) [circle,fill,inner sep=1.2pt,label=above:$u_1$] {};
\node at (4.5,1) [circle,fill,inner sep=1.2pt,label=above:$u_2$] {};
\draw (4,0) to[out=90, in=270] (4,1);
\draw (4.5,0) to[out=90, in=270] (4.5,1);
\node at (4.7,0.4) {$.$};
\end{tikzpicture}
\end{equation}
The second solution fixes the outgoing rapidities $v_{1,2}$ and relates the coefficients $A_{12}$ and $B_{12}$ via
\begin{align}
&v_1=u_2,
&
& v_2=u_1,
& A_{12}=\text S_{12}^0 \delta_{v_1,u_2}\delta_{v_2,u_1},
& 
& B_{12}=\frac{i}{u_{12}}\text S_{12}^0\delta_{v_1,u_2}\delta_{v_2,u_1},\label{sol2}
\end{align}
where $\text S_{12}^0$ is an undetermined overall factor that may depend on the free rapidities. This solution describes the nontrivial scattering of two particles as the rapidities in the final state are exchanged. In terms of the permutation diagrams this result for the two-particle S-matrix $\Smat_{12}$ can be illustrated as 
\begin{equation}
\Smat_{12}(u_{1,2};v_{1,2})=\text S_{12}^0\ \delta_{u_1,v_2}\delta_{u_2,v_1}\left[\rule{0cm}{.9cm}\right.
\begin{tikzpicture}[thick,scale=0.7, every node/.style={scale=0.7},baseline={([yshift=-.5ex]current bounding box.center)}]
\node at (3.6,0) [circle,fill,inner sep=1.2pt,label=below:$u_2$] {};
\node at (4.1,0) [circle,fill,inner sep=1.2pt,label=below:$u_1$] {};
\node at (3.6,1) [circle,fill,inner sep=1.2pt,label=above:$u_1$] {};
\node at (4.1,1) [circle,fill,inner sep=1.2pt,label=above:$u_2$] {};
\draw (3.6,0) to[out=95, in=275] (4.1,1);
\draw (4.1,0) to[out=85, in=265] (3.6,1);
\end{tikzpicture}
+\dfrac{i}{u_{12}}
\begin{tikzpicture}[thick,scale=0.7, every node/.style={scale=0.7},baseline={([yshift=-.5ex]current bounding box.center)}]
\node at (5.9,0) [circle,fill,inner sep=1.2pt,label=below:$u_2$] {};
\node at (6.4,0) [circle,fill,inner sep=1.2pt,label=below:$u_1$] {};
\node at (5.9,1) [circle,fill,inner sep=1.2pt,label=above:$u_1$] {};
\node at (6.4,1) [circle,fill,inner sep=1.2pt,label=above:$u_2$] {};
\draw (5.9,0) to[out=90, in=270] (5.9,1);
\draw (6.4,0) to[out=90, in=270] (6.4,1);
\node at (6.8,0.5) {$\left.\rule{0cm}{1.2cm}\right]$};
\end{tikzpicture}
\label{sun2}
\end{equation}
and maps as $
\text S_{12}: V_{u_2}\otimes V_{u_1}\rightarrow V_{u_1}\otimes V_{u_2}. 
$
Thus the Yangian constraint relates the coefficients $A_{12}$ and $B_{12}$ in \eqref{sunl02} via a factor depending on the difference of the particles' rapidities but, expectedly, does not fix the overall factor. The set of rapidities comes out to be conserved and we do not need to impose momentum conservation on top of the Yangian constraints. Furthermore, up to the rapidity-dependence of the overall factor, the S-matrix only depends on the rapidity differences of the incoming particles. Note that the result is independent of the rank of $\alg{su}(N)$.

\paragraph{Three-Particle S-matrix.}

Again it is useful to employ permutation symbols to parametrize the Yangian invariant. The most generic $\mathfrak{su}(N)$ (i.e.\ level-0) invariant operator of length 3 can then be written as the linear combination
\begin{align}
\mathcal I_{123}(u_i,v_i)=A_{123}
\begin{tikzpicture}[thick,scale=0.7, every node/.style={scale=0.7},baseline={([yshift=-.5ex]current bounding box.center)}]
\node at (2.5,0) [circle,fill,inner sep=1.2pt,label=below:$v_1$] {};
\node at (3.0,0) [circle,fill,inner sep=1.2pt,label=below:$v_2$] {};
\node at (3.5,0) [circle,fill,inner sep=1.2pt,label=below:$v_3$] {};
\node at (2.5,1) [circle,fill,inner sep=1.2pt,label=above:$u_1$] {};
\node at (3.0,1) [circle,fill,inner sep=1.2pt,label=above:$u_2$] {};
\node at (3.5,1) [circle,fill,inner sep=1.2pt,label=above:$u_3$] {};
\draw (2.5,0) to[out=95, in=275] (3.5,1);
\draw (3.0,0) to[out=90, in=270] (3.0,1);
\draw (3.5,0) to[out=85, in=265] (2.5,1);
\end{tikzpicture}
+B_{123}
\begin{tikzpicture}[thick,scale=0.7, every node/.style={scale=0.7},baseline={([yshift=-.5ex]current bounding box.center)}]
\node at (5.0,0) [circle,fill,inner sep=1.2pt,label=below:$v_1$] {};
\node at (5.5,0) [circle,fill,inner sep=1.2pt,label=below:$v_2$] {};
\node at (6.0,0) [circle,fill,inner sep=1.2pt,label=below:$v_3$] {};
\node at (5,1) [circle,fill,inner sep=1.2pt,label=above:$u_1$] {};
\node at (5.5,1) [circle,fill,inner sep=1.2pt,label=above:$u_2$] {};
\node at (6,1) [circle,fill,inner sep=1.2pt,label=above:$u_3$] {};
\draw (5,0) to[out=95, in=275] (5.5,1);
\draw (5.5,0) to[out=95, in=275] (6.0,1);
\draw (6.0,0) to[out=85, in=265] (5,1);
\end{tikzpicture}
+C_{123}
\begin{tikzpicture}[thick,scale=0.7, every node/.style={scale=0.7},baseline={([yshift=-.5ex]current bounding box.center)}]
\node at (7.5,0) [circle,fill,inner sep=1.2pt,label=below:$v_1$] {};
\node at (8,0) [circle,fill,inner sep=1.2pt,label=below:$v_2$] {};
\node at (8.5,0) [circle,fill,inner sep=1.2pt,label=below:$v_3$] {};
\node at (7.5,1) [circle,fill,inner sep=1.2pt,label=above:$u_1$] {};
\node at (8,1) [circle,fill,inner sep=1.2pt,label=above:$u_2$] {};
\node at (8.5,1) [circle,fill,inner sep=1.2pt,label=above:$u_3$] {};
\draw (7.5,0) to[out=95, in=275] (8.5,1);
\draw (8,0) to[out=85, in=265] (7.5,1);
\draw (8.5,0) to[out=85, in=265] (8,1);
\end{tikzpicture}
+D_{123}
\begin{tikzpicture}[thick,scale=0.7, every node/.style={scale=0.7},baseline={([yshift=-.5ex]current bounding box.center)}]
\node at (3,-2.5) [circle,fill,inner sep=1.2pt,label=below:$v_1$] {};
\node at (3.5,-2.5) [circle,fill,inner sep=1.2pt,label=below:$v_2$] {};
\node at (4,-2.5) [circle,fill,inner sep=1.2pt,label=below:$v_3$] {};
\node at (3,-1.5) [circle,fill,inner sep=1.2pt,label=above:$u_1$] {};
\node at (3.5,-1.5) [circle,fill,inner sep=1.2pt,label=above:$u_2$] {};
\node at (4,-1.5) [circle,fill,inner sep=1.2pt,label=above:$u_3$] {};
\draw (3,-2.5) to[out=90, in=270] (3,-1.5);
\draw (3.5,-2.5) to[out=95, in=275] (4,-1.5);
\draw (4,-2.5) to[out=85, in=265] (3.5,-1.5);
\end{tikzpicture}
+E_{123}
\begin{tikzpicture}[thick,scale=0.7, every node/.style={scale=0.7},baseline={([yshift=-.5ex]current bounding box.center)}]
\node at (5.5,-2.5) [circle,fill,inner sep=1.2pt,label=below:$v_1$] {};
\node at (6,-2.5) [circle,fill,inner sep=1.2pt,label=below:$v_2$] {};
\node at (6.5,-2.5) [circle,fill,inner sep=1.2pt,label=below:$v_3$] {};
\node at (5.5,-1.5) [circle,fill,inner sep=1.2pt,label=above:$u_1$] {};
\node at (6,-1.5) [circle,fill,inner sep=1.2pt,label=above:$u_2$] {};
\node at (6.5,-1.5) [circle,fill,inner sep=1.2pt,label=above:$u_3$] {};
\draw (6.5,-2.5) to[out=90, in=270] (6.5,-1.5);
\draw (5.5,-2.5) to[out=95, in=275] (6.0,-1.5);
\draw (6.0,-2.5) to[out=85, in=265] (5.5,-1.5);
\end{tikzpicture}
+F_{123}
\begin{tikzpicture}[thick,scale=0.7, every node/.style={scale=0.7},baseline={([yshift=-.5ex]current bounding box.center)}]
\node at (8,-2.5) [circle,fill,inner sep=1.2pt,label=below:$v_1$] {};
\node at (8.5,-2.5) [circle,fill,inner sep=1.2pt,label=below:$v_2$] {};
\node at (9,-2.5) [circle,fill,inner sep=1.2pt,label=below:$v_3$] {};
\node at (8,-1.5) [circle,fill,inner sep=1.2pt,label=above:$u_1$] {};
\node at (8.5,-1.5) [circle,fill,inner sep=1.2pt,label=above:$u_2$] {};
\node at (9,-1.5) [circle,fill,inner sep=1.2pt,label=above:$u_3$] {};
\draw (8,-2.5) to[out=90, in=270] (8,-1.5);
\draw (9,-2.5) to[out=90, in=270] (9,-1.5);
\draw (8.5,-2.5) to[out=90, in=270] (8.5,-1.5);
\end{tikzpicture}\label{sunl03}
\end{align}
with coefficients $A_{123},...,F_{123}$ which may be rapidity dependent.
%
For $N=2$ only five of these six diagrams are linearly independent.
Imposing the level-1 constraints fixes all coefficients up to an overall factor and yields the S-matrix $\Smat_{123}$ for three-particle scattering
\begin{align}
\Smat_{123}(u_i;v_i)
=
\text S_{123}^0\delta_{u_1,v_3}\delta_{u_2,v_2}\delta_{u_3,v_1}
\left[\rule{0cm}{.9cm}\right.
&
\begin{tikzpicture}[thick,scale=0.7, every node/.style={scale=0.7},baseline={([yshift=-.5ex]current bounding box.center)}]
\node at (3.4,0) [circle,fill,inner sep=1.2pt,label=below:$u_3$] {};
\node at (3.9,0) [circle,fill,inner sep=1.2pt,label=below:$u_2$] {};
\node at (4.4,0) [circle,fill,inner sep=1.2pt,label=below:$u_1$] {};
\node at (3.4,1) [circle,fill,inner sep=1.2pt,label=above:$u_1$] {};
\node at (3.9,1) [circle,fill,inner sep=1.2pt,label=above:$u_2$] {};
\node at (4.4,1) [circle,fill,inner sep=1.2pt,label=above:$u_3$] {};
\draw (3.4,0) to[out=95, in=275] (4.4,1);
\draw (3.9,0) to[out=90, in=270] (3.9,1);
\draw (4.4,0) to[out=85, in=265] (3.4,1);
\end{tikzpicture}
+\dfrac{i}{u_{23}}
\begin{tikzpicture}[thick,scale=0.7, every node/.style={scale=0.7},baseline={([yshift=-.5ex]current bounding box.center)}]
\node at (4.7,-2.2) [circle,fill,inner sep=1.2pt,label=below:$u_3$] {};
\node at (5.2,-2.2) [circle,fill,inner sep=1.2pt,label=below:$u_2$] {};
\node at (5.7,-2.2) [circle,fill,inner sep=1.2pt,label=below:$u_1$] {};
\node at (4.7,-1.2) [circle,fill,inner sep=1.2pt,label=above:$u_1$] {};
\node at (5.2,-1.2) [circle,fill,inner sep=1.2pt,label=above:$u_2$] {};
\node at (5.7,-1.2) [circle,fill,inner sep=1.2pt,label=above:$u_3$] {};
\draw (4.7,-2.2) to[out=95, in=275] (5.2,-1.2);
\draw (5.2,-2.2) to[out=95, in=275] (5.7,-1.2);
\draw (5.7,-2.2) to[out=85, in=265] (4.7,-1.2);
\end{tikzpicture}
+\dfrac{i}{u_{13}}
\begin{tikzpicture}[thick,scale=0.7, every node/.style={scale=0.7},baseline={([yshift=-.5ex]current bounding box.center)}]
\node at (7.5,-2.2) [circle,fill,inner sep=1.2pt,label=below:$u_3$] {};
\node at (8,-2.2) [circle,fill,inner sep=1.2pt,label=below:$u_2$] {};
\node at (8.5,-2.2) [circle,fill,inner sep=1.2pt,label=below:$u_1$] {};
\node at (7.5,-1.2) [circle,fill,inner sep=1.2pt,label=above:$u_1$] {};
\node at (8,-1.2) [circle,fill,inner sep=1.2pt,label=above:$u_2$] {};
\node at (8.5,-1.2) [circle,fill,inner sep=1.2pt,label=above:$u_3$] {};
\draw (7.5,-2.2) to[out=90, in=270] (7.5,-1.2);
\draw (8.5,-2.2) to[out=90, in=270] (8.5,-1.2);
\draw (8,-2.2) to[out=90, in=270] (8,-1.2);
\end{tikzpicture}
+\dfrac{i}{u_{12}}
\begin{tikzpicture}[thick,scale=0.7, every node/.style={scale=0.7},baseline={([yshift=-.5ex]current bounding box.center)}]
\node at (10.3,-2.2) [circle,fill,inner sep=1.2pt,label=below:$u_3$] {};
\node at (10.8,-2.2) [circle,fill,inner sep=1.2pt,label=below:$u_2$] {};
\node at (11.3,-2.2) [circle,fill,inner sep=1.2pt,label=below:$u_1$] {};
\node at (10.3,-1.2) [circle,fill,inner sep=1.2pt,label=above:$u_1$] {};
\node at (10.8,-1.2) [circle,fill,inner sep=1.2pt,label=above:$u_2$] {};
\node at (11.3,-1.2) [circle,fill,inner sep=1.2pt,label=above:$u_3$] {};
\draw (10.3,-2.2) to[out=95, in=275] (11.3,-1.2);
\draw (10.8,-2.2) to[out=85, in=265] (10.3,-1.2);
\draw (11.3,-2.2) to[out=85, in=265] (10.8,-1.2);
\end{tikzpicture}
\nonumber\\
&
-\dfrac{1}{u_{12} u_{23}}\left[\rule{0cm}{.9cm}\right.
\begin{tikzpicture}[thick,scale=0.7, every node/.style={scale=0.7},baseline={([yshift=-.5ex]current bounding box.center)}]
\node at (8,-4.4) [circle,fill,inner sep=1.2pt,label=below:$u_3$] {};
\node at (8.5,-4.4) [circle,fill,inner sep=1.2pt,label=below:$u_2$] {};
\node at (9,-4.4) [circle,fill,inner sep=1.2pt,label=below:$u_1$] {};
\node at (8,-3.4) [circle,fill,inner sep=1.2pt,label=above:$u_1$] {};
\node at (8.5,-3.4) [circle,fill,inner sep=1.2pt,label=above:$u_2$] {};
\node at (9,-3.4) [circle,fill,inner sep=1.2pt,label=above:$u_3$] {};
\draw (8,-4.4) to[out=90, in=270] (8,-3.4);
\draw (8.5,-4.4) to[out=95, in=275] (9,-3.4);
\draw (9,-4.4) to[out=85, in=265] (8.5,-3.4);
\end{tikzpicture}
+
\begin{tikzpicture}[thick,scale=0.7, every node/.style={scale=0.7},baseline={([yshift=-.5ex]current bounding box.center)}]
\node at (10,-4.4) [circle,fill,inner sep=1.2pt,label=below:$u_3$] {};
\node at (10.5,-4.4) [circle,fill,inner sep=1.2pt,label=below:$u_2$] {};
\node at (11,-4.4) [circle,fill,inner sep=1.2pt,label=below:$u_1$] {};
\node at (10,-3.4) [circle,fill,inner sep=1.2pt,label=above:$u_1$] {};
\node at (10.5,-3.4) [circle,fill,inner sep=1.2pt,label=above:$u_2$] {};
\node at (11,-3.4) [circle,fill,inner sep=1.2pt,label=above:$u_3$] {};
\draw (11,-4.4) to[out=90, in=270] (11,-3.4);
\draw (10,-4.4) to[out=95, in=275] (10.5,-3.4);
\draw (10.5,-4.4) to[out=85, in=265] (10,-3.4);
\node at (11.4,-3.9) {$\left.\rule{0cm}{1.2cm}\right]$};
\end{tikzpicture}
-\dfrac{i}{u_{12}u_{13}u_{23}}
\begin{tikzpicture}[thick,scale=0.7, every node/.style={scale=0.7},baseline={([yshift=-.5ex]current bounding box.center)}]
\node at (10.6,-6.6) [circle,fill,inner sep=1.2pt,label=below:$u_3$] {};
\node at (11.1,-6.6) [circle,fill,inner sep=1.2pt,label=below:$u_2$] {};
\node at (11.6,-6.6) [circle,fill,inner sep=1.2pt,label=below:$u_1$] {};
\node at (10.6,-5.6) [circle,fill,inner sep=1.2pt,label=above:$u_1$] {};
\node at (11.1,-5.6) [circle,fill,inner sep=1.2pt,label=above:$u_2$] {};
\node at (11.6,-5.6) [circle,fill,inner sep=1.2pt,label=above:$u_3$] {};
\draw (10.6,-6.6) to[out=90, in=270] (10.6,-5.6);
\draw (11.6,-6.6) to[out=90, in=270] (11.6,-5.6);
\draw (11.1,-6.6) to[out=90, in=270] (11.1,-5.6);
\end{tikzpicture}
\left.\rule{0cm}{.9cm}\right].
\label{sun3}
\end{align}
Note that the constraint equations allow for five more solutions which are incompatible with the condition $u_1>u_2>u_3$ and $v_1<v_2<v_3$ for three-particle scattering.

The explicit form \eqref{sun2, sun3} of the two- and three-particle S-matrices at hand, we  have checked consistent factorization by verifying the equalities \eqref{3fact} and the qYBE \eqref{qYBE}.
This shows that the unique Yangian-invariant three-particle S-matrix is indeed given by the factorized form.

To summarize, Yangian invariance completely fixes the two- and three-particle scattering matrices up to an overall factor. Generically, the overall scattering phase can be determined by imposing additional unitarity and crossing relations, see e.g.\ \cite{Bombardelli:2016scq}. 
Notably, Yangian symmetry restricts the set of outgoing momenta to be the same as the set of incoming momenta, i.e.\
\begin{align}
\{u_i\}=\{v_i\}.
\end{align}
 This is a substantial feature of scattering processes in integrable (1+1)-dimensional models which automatically follows from the Yangian constraints here.
Moreover, the S-matrices only depend on differences $u_{ij}$ of the rapidities as expected from the discussion in \secref{sec:diffform}. The three-particle S-matrix factorizes consistently into two-particle S-matrices.

\subsection{Yangian Constraints for $\alg{u}(1|1)$}
\label{subsec:u11}

We now move on to the analysis of the Yangian constraints in the evaluation representation for the fundamental representation of Lie superalgebras. As a first step towards the algebra $\mathfrak{su}(2|2)$ we consider $\mathfrak{u}(1|1)$, i.e.\ the model discussed in \cite{Beisert:2005wm}.


\paragraph{$Y[\mathfrak{u}(1|1)]$ in the Evaluation Representation.} 
The level-0 algebra of $Y[\mathfrak{u}(1|1)]$ consists of the supersymmetry generators $\mathfrak Q$ and $\mathfrak S$, the central charge $\mathfrak C$ and the generator $\mathfrak B$ which is an outer automorphism of $\mathfrak{su}(1|1)$. These generators satisfy the commutation relations
\begin{align}
&\{\mathfrak Q,\mathfrak S\}=\mathfrak C,& &[\mathfrak B,\mathfrak Q]=-2\mathfrak Q,& &[\mathfrak B,\mathfrak S]=+2\mathfrak S.&
\end{align}
The remaining Lie brackets vanish.
The fundamental representation of this algebra consists of one bosonic state $\ket{\phi}$ and one fermionic state $\ket{\psi}$. The generators of $\mathfrak{u}(1|1)$ act on these single-particle states as
\begin{align}
&\mathfrak Q\ket{\phi}=q\ket{\psi},& &\mathfrak Q\ket{\psi}=0,&\nonumber\\
&\mathfrak S\ket{\phi}=0,& &\mathfrak S\ket{\psi}=\tfrac cq\ket{\phi},&\nonumber\\
&\mathfrak B\ket{\phi}=(b+1)\ket{\phi},& &\mathfrak B\ket{\psi}=(b-1)\ket{\psi},&\nonumber\\
&\mathfrak C\ket{\phi}=c\ket{\phi},& &\mathfrak C\ket{\psi}=c\ket{\psi},&
\label{su11fb}
\end{align}
with parameters $b$, $c$ and $q$. Note that $q$ corresponds to an unphysical rescaling of $\phi$ and $\psi$.

The evaluation representation of the level-1 generators $\widehat{\mathfrak Q},\ \widehat{\mathfrak S},\ \widehat{\mathfrak B},\ \widehat{\mathfrak C}$ (collectively denoted by $\levo$) again lifts the Lie algebra representation via
\begin{align}
\levo_i\ket{\Phi,u}=igu\levz_i\ket{\Phi,u},
\label{eq:evrepsuperalgebra}
\end{align}
with $\Phi\in\{\phi,\psi\}$, evaluation parameter $u$ and coupling constant $g$.

We consider this evaluation representation of $Y[\mathfrak{u}(1|1)]$ both in the conventional undynamic setup and the more exotic dynamic setup in the sense discussed in \secref{subsecDynUndyn}. For the undynamic case the eigenvalues of the symmetry generators are constant, i.e.\ $b,\ c$ and $q$ are rapidity-independent and there will be no braiding factor in the coproduct of the Yangian generators. In the dynamic case the coefficients in \eqref{su11fb} become functions of the rapidity, i.e.\ $c=c(u),b=b(u),q=q(u)$. These functions are model-dependent and we do not fix their explicit form. Writing down the representation of an $n$-particle asymptotic state in the dynamic representation, we label each one-particle representation by these unrestricted functions, i.e.\ as
\begin{align}
(1|1)_{u_1,c(u_1),b(u_1),q(u_1)}\otimes(1|1)_{u_2,c(u_2),b(u_2),q(u_2)}\otimes...\otimes(1|1)_{u_n,c(u_n),b(u_n),q(u_n)},\label{Hilbertsu11}
\end{align}
 in order to make this dependence more apparent in our notation. Again the S-matrix is a map between these states.


\paragraph{$Y[\mathfrak{u}(1|1)]$-Constraints on the S-Matrix.}

In order to explicitly evaluate the level-0 constraint of the two- and three-particle S-matrix, we give the coproduct structure of $Y[\mathfrak{u}(1|1)]$ for the level-0 generators
\begin{align}
&\Delta\mathfrak Q=\mathfrak Q\otimes 1+(-1)^F \otimes \mathfrak Q,& &\Delta\mathfrak S=\mathfrak S\otimes 1+(-1)^F \otimes \mathfrak S,&\nonumber\\
&\Delta\mathfrak B=\mathfrak B\otimes 1+ 1\otimes \mathfrak B,& &\Delta\mathfrak C=\mathfrak C\otimes 1+ 1\otimes \mathfrak C,&\label{coproductsu11l0}
\end{align}
and level-1 generators\footnote{We obtained the coproduct structure of the level-1 generators from Table 2 of \cite{Beisert:2007ds}. This table shows the coproduct of the level-1 generators of $\mathfrak{su}(2|2)$. See the end of section \ref{sec:su22} for a discussion of how to obtain the $\mathfrak{u}(1|1)$ generators from the $\mathfrak{su}(2|2)$ generators.}
\begin{align}
&\Delta\widehat{\mathfrak Q}=\widehat{\mathfrak Q}\otimes 1+(-1)^F \otimes \widehat{\mathfrak Q}+\tfrac 12\mathfrak Q\otimes \mathfrak C-\tfrac 12(-1)^F\mathfrak C\otimes \mathfrak Q,\nonumber\\
&\Delta\widehat{\mathfrak S}=\widehat{\mathfrak S}\otimes 1+(-1)^F \otimes \widehat{\mathfrak S}-\tfrac 12 \mathfrak S\otimes \mathfrak C+\tfrac 12(-1)^F\mathfrak C\otimes\mathfrak S,\nonumber\\
&\Delta\widehat{\mathfrak B}=\widehat{\mathfrak B}\otimes 1+ 1\otimes \widehat{\mathfrak B}-(-1)^F\mathfrak S\otimes\mathfrak Q-(-1)^F\mathfrak Q\otimes \mathfrak S,\nonumber\\
&\Delta\widehat{\mathfrak C}=\widehat{\mathfrak C}\otimes 1+ 1\otimes \widehat{\mathfrak C}.\label{coproductsu11l1}
\end{align}
In these formulae we took into account the fermionic nature of the generators $\mathfrak Q$ and $\mathfrak S$ and their level-1 versions. It results in factors $(-1)^F$ whenever they pass through a particle. For bosonic particles we take $F=0$ and for fermionic particles $F=1$, i.e.
\begin{align}
&(-1)^F\ket{\phi}=+\ket{\phi},
&
& (-1)^F\ket{\psi}=-\ket{\psi}.
\end{align}
For the action on states including more particles this coproduct structure can be easily generalized. Note that the coproduct structure of both the undynamic and the dynamic case do not contain a braiding factor.


\paragraph{Two-Particle Undynamic S-Matrix.}

We will now present the results of the analysis of the constraints on two-particle S-matrices  \eqref{level0} and \eqref{level1} using \eqref{coproductsu11l0} and \eqref{coproductsu11l1} in the undynamic setup. In order to do so, we make an ansatz for the S-matrix by writing down a linear combination of all possible maps between asymptotic states of the form \eqref{asym} with $n=2$, i.e. 
\begin{align}
\Smat_{12}(u_{1,2};v_{1,2}): (1|1)_{v_1}\otimes(1|1)_{v_2}\rightarrow (1|1)_{u_1}\otimes(1|1)_{u_2}.
\end{align}
The level-0 constraint arising from the generator $\mathfrak B$ implies that this gets reduced to
\begin{align}
&\text S_{12}(u_i,v_i)\ket{\phi,v_1;\phi,v_2}=A_{12}\ket{\phi,u_1;\phi,u_2},\nonumber\\
&\text S_{12}(u_i,v_i)\ket{\phi,v_1;\psi,v_2}=B_{12}\ket{\psi,u_1;\phi,u_2}+C_{12}\ket{\phi,u_1;\psi,u_2},\nonumber\\
&\text S_{12}(u_i,v_i)\ket{\psi,v_1;\phi,v_2}=D_{12}\ket{\phi,u_1;\psi,u_2}+E_{12}\ket{\psi,u_1;\phi,u_2},\nonumber\\
&\text S_{12}(u_i,v_i)\ket{\psi,v_1;\psi,v_2}=F_{12}\ket{\psi,u_1;\psi,u_2},\label{Smatrixsu11}
\end{align}
with six undetermined coefficients $A_{12},...,F_{12}$. Demanding also the remaining level-0 and level-1 constraints implies that the S-matrix preserves the set of rapidities,\footnote{In the following, we only discuss the solutions of the constraint equations that are compatible with true scattering solutions, i.e.\ $u_1>u_2(>u_3)$ and $v_1<v_2(<v_3)$ for the two(three)-particle scattering process.} i.e.
\begin{align}
&v_1=u_2,& &v_2=u_1.
\label{rap2conserv}
\end{align}
Furthermore, the six coefficients in \eqref{Smatrixsu11} are related via\footnote{Note that we drop the factor of $\delta_{v_1,u_2}\delta_{v_2,u_1}$ here. For simplicity of the notation, we will continue doing so in the rest of this chapter whenever we have indicated that for the case under consideration (including three-particle scattering with factors $\delta_{v_1,u_3}\delta_{v_2,u_2}\delta_{v_3,u_1}$) the sets of rapidities are conserved.}
\begin{align}
&A_{12}=-\text S_{12}^0\frac{\tfrac{c}{ig}-u_{12}}{\tfrac{c}{ig}+u_{12}},
&
&B_{12}=D_{12}=\text S_{12}^0\frac{u_{12}}{\tfrac{c}{ig}+u_{12}},
&
&C_{12}=E_{12}=-\text S_{12}^0\frac{\tfrac{c}{ig}}{\tfrac{c}{ig}+u_{12}},
&
&F_{12}=-\text S_{12}^0.
\label{eq:S11}
\end{align}
Here $\Smat_{12}^0$ denotes an unknown overall constant that has to be determined via unitarity and crossing relations. 
Up to the rapidity dependence of the overall factor, this two-particle S-matrix only depends on the difference of the two incoming rapidities as expected from the arguments above \eqref{eq:diffform}. Furthermore this solution satisfies the quantum Yang--Baxter equation~\eqref{qYBE}.


\paragraph{Three-Particle Undynamic S-Matrix.}

We now move on to a similar analysis of the $Y[\mathfrak{u}(1|1)]$ constraints for the three-particle case. The S-matrix is the map
\begin{align}
&\text S_{123}(u_i;v_i):(1|1)_{v_1}\otimes(1|1)_{v_2}\otimes(1|1)_{v_3}\rightarrow (1|1)_{u_1}\otimes(1|1)_{u_2}\otimes(1|1)_{u_3}.
\end{align}
Again the level-0 constraint associated to $\mathfrak B$ implies that S$_{123}$ only permutes bosons and fermions and thus is a linear combination of 20 terms with unknown coefficients. The number of degrees of freedom gets reduced to 6 by imposing the remaining level-0 constraints. The level-1 constraints then impose constraints on the remaining six coefficients of the S-matrix and the incoming rapidities $v_i$. Looking for all of the solutions is rather complicated even in \texttt{Mathematica} because the set of equations contains very long expressions. Therefore we only solved it for the cases in which the representation coefficients take the exemplary values $c=q=1$ with $b=0,1,2$. Restricting to these cases we find that the outgoing rapidities have to satisfy
\begin{align}
&v_1=u_3,& 
&v_2=u_2,& 
&v_3=u_1.
\label{conserved3}
\end{align}
One can show that \eqref{conserved3} solves the constraint equations for general $c,\ q$ and $b$. In this case the level-1 constraint can also be used to fix the S-matrix up to an overall factor. It factorizes consistently into three two-particle S-matrices from above via \eqref{3fact}.


\paragraph{Two-Particle Dynamic S-Matrix.}

The dynamic two-to-two-particle S-matrix maps between asymptotic states of the form \eqref{Hilbertsu11} with $n=2$, i.e.
\begin{align}
\Smat_{12}: (1|1)_{v_1,c(v_1),b(v_1),q(v_1)}\otimes(1|1)_{v_2,c(v_2),b(v_2),q(v_2)}\rightarrow (1|1)_{u_1,c(u_1),b(u_1),q(u_1)}\otimes(1|1)_{u_2,c(u_2),b(u_2),q(u_2)}.
\end{align}
Writing down the most general ansatz of this form and constraining it via demanding its invariance under $\mathfrak B$ implies that it takes the form \eqref{Smatrixsu11} as in the undynamic case. Again its coefficients and the rapidities get further constrained by the remaining level-0 and -1 constraints. We verified that these are solved by a conserved set of rapidities \eqref{rap2conserv} with further restrictions on the S-matrix. 
It is convenient to express these in terms of the variables $x_{i}^\pm$ related to the rapidities $u_i$ and central charge eigenvalue $c$ via \cite{Beisert:2005wm}
\begin{align}
&u_i=\tfrac g2(x_i^++x_i^-),
& 
&c(u_i)=-ig(x_i^+-x_i^-).
\label{eq:xpxm}
\end{align}
Then, the constraints fix the six coefficients in \eqref{Smatrixsu11} up to an overall factor $\Smat_{12}^0$. We find them to be given by
\begin{align}
&A_{12}=\text S_{12}^0\frac{x_1^+-x_2^-}{x_1^--x_2^+},
&
&B_{12}=\text S_{12}^0\frac{x_1^+-x_2^+}{x_1^--x_2^+},
&
&C_{12}=\text S_{12}^0\frac{q_2}{q_1}\frac{x_1^+-x_1^-}{x_1^--x_2^+},\nonumber\\
&D_{12}=\text S_{12}^0\frac{x_1^--x_2^-}{x_1^--x_2^+},
&
&E_{12}=\text S_{12}^0\frac{q_1}{q_2}\frac{x_2^+-x_2^-}{x_1^--x_2^+},
&
& F_{12}=-\text S_{12}^0.\label{coefssu11}
\end{align}

The question arises whether the conservation of the set of rapidities is the unique solution to the constraint equations. Since we could not solve the involved constraints for a generic central charge eigenvalue, as an example we further specify the eigenvalue function $c(p)$ to be the energy function of the $\mathcal N=4$ SYM spin chain:
\begin{align}
c(p)&= \tfrac{1}{2}
\sqrt{1+16 g^2 \sin^2 \tfrac{p}{2}}+\tfrac 12,
\label{eq:N4energy}
\end{align}
where $g$ corresponds to the \sym\ theory coupling.
We then employ momentum conservation and the conservation of the central charge in the above form, namely
\begin{align}
p_1+p_2&=q_1+q_2,
&
c(p_1)+c(p_2)&=c(q_1)+c(q_2).
\label{eq:pEconservation}
\end{align}
Using random real values for the incoming momenta $p_{1,2}$ and the coupling $g$, the conservation equations \eqref{eq:pEconservation} can then be solved numerically for the outgoing momenta $q_{1,2}$ (related to the rapidities $u_{1,2}$ and $v_{1,2}$, respectively). This confirms that for the exemplary function \eqref{eq:N4energy} the set of incoming and outgoing momenta are indeed the same. 

The above solution represents the well-known two-particle S-matrix of the dynamic $\mathfrak{u}(1|1)$ spin chain of \cite{Beisert:2005wm}. It satisfies the quantum Yang--Baxter equation \eqref{qYBE}. For $c,b,q=\mathrm{const.}$ it simplifies to the undynamic two-particle S-matrix given in \eqref{eq:S11} via $x^\pm=u\mp \tfrac{c}{2ig}$.


\paragraph{Three-Particle Dynamic S-Matrix.}

We conclude the analysis of the $Y[\mathfrak{u}(1|1)]$ constraints by looking at the three-particle dynamic S-matrix which maps as
\begin{align}
&\text S_{123}(u_i;v_i):(1|1)_{v_1,c(v_1),b(v_1),q(v_1)}\otimes(1|1)_{v_2,c(v_2),b(v_2),q(v_2)}\otimes(1|1)_{v_3,c(v_3),b(v_3),q(v_3)}\nonumber\\
&\hspace{2.5cm}\rightarrow (1|1)_{u_1,c(u_1),b(u_1),q(u_1)}\otimes(1|1)_{u_2,c(u_2),b(u_2),q(u_2)}\otimes(1|1)_{u_3,c(u_3),b(u_3),q(u_3)}.
\end{align}
The level-0 constraint associated to $\mathfrak B$ implies as in the undynamic case that S$_{123}$ is a linear combination with 20 unknown coefficients. 
For generic functions $c(u)$, $b(u)$ and $q(u)$ and different sets of incoming and outgoing rapidities, the remaining Yangian constraint equations were not tractable in \texttt{Mathematica}; neither could we solve three-particle generalizations of equations \eqref{eq:pEconservation}. However, we verified that the remaining constraints are solved by conserved sets of rapidities \eqref{conserved3}. 
Given these conserved sets of rapidities, we have then verified that the unique three-particle solution to the Yangian constraints is given by the factorized S-matrix of the form \eqref{eq:S11,coefssu11}. 

Let us summarize the results of this subsection. The Yangian constraints on S-matrices impose conserved sets of rapidities in two- and three-particle scattering events in the undynamic case. Also in the dynamic setup these constraints are solved by conserved sets of rapidities, but we could not verify the uniqueness of this solution in the three-particle case. Specifying the central charge eigenvalue to the AdS/CFT energy-momentum relation \eqref{eq:N4energy}, in the two-particle case we have argued for its uniqueness from momentum and energy conservation. 
For conserved sets of rapidities, the two- and three-particle undynamic and dynamic $Y[\mathfrak{u}(1|1)]$-invariant S-matrix is uniquely determined up to the overall dressing factor. The three-particle S-matrices factorize consistently into three two-particle S-matrices. As expected by the discussion in \secref{sec:diffform}, the resulting undynamic S-matrices are of difference form, while the dynamic ones are not.


\subsection{Yangian Constraints for $\alg{su}(2|2)(\ltimes \mathbb{R}^2)$}
\label{sec:su22}

The third Yangian algebra we investigated with respect to its constraints on S-matrices is the Yangian corresponding to $\mathfrak{su}(2|2)$ and its central extension $\mathfrak{su}(2|2)\ltimes\mathbb R^2$, respectively.


\paragraph{$Y[\mathfrak{su}(2|2)]$ and $Y[\mathfrak{su}(2|2)\ltimes\mathbb R^2]$ in the Evaluation Representation.} 

We begin with a discussion of the Lie superalgebra $\mathfrak{su}$(2|2) on the basis of \cite{Beisert:2006qh,Beisert:2007ds}. It consists of the $\mathfrak{su}$(2)$\times\mathfrak{su}$(2) generators $\mathfrak R^a_b$ and $\mathfrak L^\alpha_\beta$, the supersymmetry generators $\mathfrak Q^\alpha_b$, $\mathfrak S^a_\beta$ and the central charge $\mathfrak C$. Note that we use Latin indices $a,b,c,...$ to denote bosonic degrees of freedom and Greek indices $\alpha,\beta,\gamma,...$ for fermionic degrees of freedom. These generators satisfy the commutation relations
\begin{align}
&[\mathfrak R^a_b,\mathfrak R^c_d]=\delta_b^c\mathfrak R^a_d-\delta^a_d\mathfrak R^c_b, \hspace{2.05cm} [\mathfrak L^\alpha_\beta,\mathfrak L^\gamma_\delta]=\delta_\beta^\gamma\mathfrak L^\alpha_\delta-\delta^\alpha_\delta\mathfrak L^\gamma_\beta,\nonumber\\
&[\mathfrak R^a_b,\mathfrak Q^\gamma_d]=-\delta^a_d\mathfrak Q^\gamma_b+\tfrac 12\delta^a_b\mathfrak Q^\gamma_d,\hspace{1.4cm}[\mathfrak L^\alpha_\beta,\mathfrak Q^\gamma_d]=+\delta^\gamma_\beta\mathfrak Q^\alpha_d-\tfrac 12\delta^\alpha_\beta\mathfrak Q^\gamma_d,\nonumber\\
&[\mathfrak R^a_b,\mathfrak S^c_\delta]=+\delta^c_b\mathfrak S^a_\delta-\tfrac 12\delta^a_b\mathfrak S^c_\delta,\hspace{1.5cm}[\mathfrak L^\alpha_\beta,\mathfrak S^c_\delta]=-\delta^\alpha_\delta\mathfrak S^c_\beta+\tfrac 12\delta^\alpha_\beta\mathfrak S^c_\delta,\nonumber\\
&\{\mathfrak Q^\alpha_b,\mathfrak S^c_\delta\}=\delta^c_b\mathfrak L^\alpha_\delta+\delta^\alpha_\delta\mathfrak R^c_b+\delta^c_b\delta^\alpha_\delta\mathfrak C.\label{algebraRLC}
\end{align}
The remaining Lie brackets vanish. In order to obtain the correct S-matrix for the planar limit of $\mathcal N=4$ SYM we extend this algebra by two central charges $\mathfrak P$ and $\mathfrak K$ to $\mathfrak{su}(2|2)\ltimes\mathbb R^2$ \cite{Beisert:2006qh}. These modify the anticommutators to
\begin{align}
&\{\mathfrak Q^\alpha_b,\mathfrak Q^\gamma_d\}=\varepsilon^{\alpha\gamma}\varepsilon_{bd}\mathfrak P\nonumber\\
&\{\mathfrak S^a_\beta,\mathfrak S^c_\delta\}=\varepsilon^{ac}\varepsilon_{\beta\delta}\mathfrak K.\label{algebraPK}
\end{align}
Demanding that (the total) $\mathfrak P$ and $\mathfrak K$ annihilate physical states constrains the eigenvalues of these operators and guarantees that the physical result is compatible with an $\mathfrak{su}$(2|2) invariance of the model. 

The fundamental representation of this algebra acts on two bosonic states $\ket{\phi^a}$ and two fermionic states $\ket{\psi^\alpha}$ with $a,\alpha=1,2$. The $\mathfrak{su}(2)$ generators $\mathfrak R^a_b$ and $\mathfrak L^\alpha_\beta$ act analogously to \eqref{sugenerators} as
\begin{align}
&\mathfrak R^a_b\ket{\phi^c}=\delta_b^c\ket{\phi^a}-\tfrac 12\delta^a_b\ket{\phi^c},
&
&\mathfrak L^\alpha_\beta\ket{\psi^\gamma}=\delta_\beta^\gamma\ket{\psi^\alpha}-\tfrac 12\delta^\alpha_\beta\ket{\psi^\gamma}. \label{su22repRL}
\end{align}
In this representation the supersymmetry generators transform bosons into fermions and vice versa with coefficients $a,b,c,d$ as 
\begin{align}
&\mathfrak Q^\alpha_a\ket{\phi^b}=a\ \delta_a^b\ket{\psi^\alpha},\qquad\qquad \mathfrak Q^\alpha_a\ket{\psi^\beta}=b\ \varepsilon^{\alpha\beta}\varepsilon_{ab}\ket{\phi^b},\nonumber\\
&\mathfrak S^a_\alpha\ket{\phi^b}=c\ \varepsilon^{ab}\varepsilon_{\alpha\beta}\ket{\psi^\beta},\qquad\ \mathfrak S^a_\alpha\ket{\psi^\beta}=d\ \delta_\alpha^\beta\ket{\phi^a}.\label{su22repQS}
\end{align}
Via the anticommutation relations \eqref{algebraPK}, the eigenvalues $P$ and $K$ of the central charges $\mathfrak P$ and $\mathfrak K$ are given by
\begin{align}
&P=ab,
& 
&K=cd.
\label{PK}
\end{align}
The anticommutator $\{\mathfrak Q^\alpha_b,\mathfrak S^c_\delta\}$ in \eqref{algebraRLC} implies that the eigenvalue $C$ of $\mathfrak C$ is given by
\begin{align}
C=\tfrac 12(ad+bc),
\label{Cad}
\end{align}
and that the coefficients $a,b,c,d$ satisfy the constraint
\begin{align}
ad-bc=1.\label{adbc}
\end{align}

\paragraph{Undynamic limit.}
Removing the central extensions $\mathfrak P$ and $\mathfrak K$ from the algebra $\mathfrak{su}(2|2)\ltimes\mathbb R^2$ corresponds to setting their eigenvalues to $0$. This implies via \eqref{Cad} and \eqref{adbc} that $C=\pm\tfrac 12$. The resulting algebra is relevant in the context of strongly correlated electron systems on a one-dimensional lattice, see \cite{Essler:1992py,Essler:1992bn,Essler:1992uc}. In this model each site is a superposition of four possible electronic states and two of these are fermionic. There are no length-changing effects such that the undynamic representation with $\mathcal U=1$ becomes relevant and one may directly put $P=K=0$ at each site with rapidity-independent coefficients $a,b,c,d$.

\paragraph{Dynamic case.}
In the $\mathfrak{su}(2|3)$ sector of $\mathcal N=4$ SYM the algebra $\mathfrak{su}(2|2)$ appears as residual symmetry algebra of the excitations of the spin chain. In this sector the central charge $\mathfrak C$ is associated to the integrable spin chain Hamiltonian and its eigenvalues are not constant as discussed above. This contradiction is resolved by the introduction of the additional generators $\mathfrak P$ and $\mathfrak K$ whose eigenvalues only have to vanish on physical states in order to guarantee the overall $\mathfrak{su}(2|2)$ symmetry of the excitations. This setup implies that the one-magnon energy is given by \cite{Beisert:2005tm} 
\begin{align}
C=\pm\tfrac 12\sqrt{1+16\alpha\beta\sin^2\left(\tfrac p2\right)},
\label{eq:Calphabeta}
\end{align}
where $\alpha$ and $\beta$ are constants satisfying
\begin{align}
&ab=g\alpha\left(1-e^{ip}\right),
& 
&cd=\tfrac{\beta}{g}\left(1-e^{-ip}\right).
\end{align}
The product $\alpha\beta=g^2$ corresponds to the square of the coupling constant $g$ of the model, i.e.\ $C=\pm\frac 12$ holds at leading order.

The Yangian corresponding to $\mathfrak{su}(2|2)\ltimes\mathbb R^2$ consists of the level-0 generators $\mathfrak R^a_b$, $\mathfrak L^\alpha_\beta$, $\mathfrak Q^\alpha_b$, $\mathfrak S^a_\beta$ and $\mathfrak C$, as well as the central charges $\mathfrak P$ and $\mathfrak K$ of the central extension. The corresponding level-1 generators $\levo$ are $\widehat{\mathfrak R}^a_b$, $\widehat{\mathfrak L}^\alpha_\beta$, $\widehat{\mathfrak Q}^\alpha_b$, $\widehat{\mathfrak S}^a_\beta, \widehat{\mathfrak C},\widehat{\mathfrak P}$ and $\widehat{\mathfrak K}$. Their action on single particles in this representation is given by \eqref{eq:evrepsuperalgebra} with $\Phi\in\{\phi^a,\psi^\alpha\}$.


\paragraph{$Y[\mathfrak{su}(2|2)]$- and $Y[\mathfrak{su}(2|2)\ltimes\mathbb R^2]$-Constraints on the S-Matrix.}

In order to evaluate explicitly the constraints \eqref{level0} and \eqref{level1}, we need the coproduct structure of the Yangian generators. For the level-0 generators acting on two sites it is \cite{Beisert:2007ds}
\begin{align}
&\Delta\mathfrak C=\mathfrak C\otimes 1+1\otimes\mathfrak C,& &\Delta\mathfrak R^a_b=\mathfrak R^a_b\otimes 1+1\otimes\mathfrak R^a_b,\nonumber\\
&\Delta\mathfrak P=\mathfrak P\otimes 1+\mathcal U^{+2}\otimes\mathfrak P,& &\Delta\mathfrak L^\alpha_\beta=\mathfrak L^\alpha_\beta\otimes 1+1\otimes\mathfrak L^\alpha_\beta,\nonumber\\
&\Delta\mathfrak K=\mathfrak K\otimes 1+\mathcal U^{-2}\otimes\mathfrak K,& &\Delta\mathfrak Q^\alpha_b=\mathfrak Q^\alpha_b\otimes 1+\mathcal U^{+1}_F\otimes\mathfrak Q^\alpha_b,
\nonumber\\
& & &\Delta\mathfrak S^a_\beta=\mathfrak S^a_\beta\otimes 1+\mathcal U^{-1}_F\otimes\mathfrak S^a_\beta\label{Y222l0}
\end{align}
with $\mathcal U_F:=(-1)^F\ \mathcal U$. 
Here $(-1)^F$ again denotes the fermionic grading operator, i.e.
\begin{align}
&(-1)^F\ket{\phi^a}=+\ket{\phi^a},
& 
&(-1)^F\ket{\psi^\alpha}=-\ket{\psi^\alpha},\label{ONE}
\end{align}
which takes care of the correct statistics when anticommuting fermionic particles and the fermionic operators $\mathfrak Q^\alpha_b$ and $\mathfrak S^a_\beta$. The abelian braiding operator $\mathcal U$ includes length-changing effects and can be set to 1 for conventional undynamic spin chains. The coproduct structure of the level-1 generators is given by
\begin{align}
\Delta\widehat{\mathfrak C}=&\widehat{\mathfrak C}\otimes 1+1\otimes\widehat{\mathfrak C}+\tfrac 12\mathfrak P\ \mathcal U^{-2}\otimes\mathfrak K-\tfrac 12\mathfrak K\ \mathcal U^{+2}\otimes\mathfrak P,\nonumber\\
\Delta\widehat{\mathfrak P}=&\widehat{\mathfrak P}\otimes 1+\mathcal U^{+2}\otimes\widehat{\mathfrak P}-\mathfrak C\ \mathcal U^{+2}\otimes\mathfrak P+\mathfrak P\otimes\mathfrak C,\nonumber\\
\Delta\widehat{\mathfrak K}=&\widehat{\mathfrak K}\otimes 1+\mathcal U^{-2}\ \otimes\widehat{\mathfrak K}+\mathfrak C\ \mathcal U^{-2}\otimes\mathfrak K-\mathfrak K\otimes\mathfrak C,\nonumber\\
\Delta\widehat{\mathfrak R}^a_b=&\widehat{\mathfrak R}^a_b\otimes 1+1\otimes \widehat{\mathfrak R}^a_b+\tfrac 12 \mathfrak R^a_c\otimes \mathfrak R^c_b-\tfrac 12\mathfrak R^c_b\otimes\mathfrak R^a_c\nonumber\\
&\ -\tfrac 12\mathfrak S^a_\gamma\ \mathcal U^{+1}_F\otimes\mathfrak Q^\gamma_b-\tfrac 12\mathfrak Q^\gamma_b\ \mathcal U^{-1}_F\otimes\mathfrak S^a_\gamma\nonumber\\
&\ +\tfrac 14\delta^a_b\ \mathfrak S^d_\gamma\ \mathcal U^{+1}_F\otimes\mathfrak Q^\gamma_d+\tfrac 14\delta^a_b\ \mathfrak Q^\gamma_d\ \mathcal U^{-1}_F\otimes\mathfrak S^d_\gamma,\nonumber\\
\Delta\widehat{\mathfrak L}^\alpha_\beta=&\widehat{\mathfrak L}^\alpha_\beta\otimes 1+1\otimes \widehat{\mathfrak L}^\alpha_\beta-\tfrac 12 \mathfrak L^\alpha_\gamma\otimes \mathfrak L^\gamma_\beta+\tfrac 12\mathfrak L^\gamma_\beta\otimes\mathfrak L^\alpha_\gamma\nonumber\\
&\ +\tfrac 12\mathfrak Q^\alpha_c\ \mathcal U^{-1}_F\otimes\mathfrak S^c_\beta+\tfrac 12\mathfrak S^c_\beta\ \mathcal U^{+1}_F\otimes\mathfrak Q^\alpha_c\nonumber\\
&\ -\tfrac 14\delta^\alpha_\beta\ \mathfrak Q^\delta_c\ \mathcal U^{-1}_F\otimes\mathfrak S^c_\delta-\tfrac 14\delta^\alpha_\beta\ \mathfrak S^c_\delta\ \mathcal U^{+1}_F\otimes\mathfrak Q^\delta_c,\nonumber\\
\Delta\widehat{\mathfrak Q}^\alpha_b=&\widehat{\mathfrak Q}^\alpha_b\otimes 1+\mathcal U^{+1}_F\otimes \widehat{\mathfrak Q}^\alpha_b-\tfrac 12 \mathfrak L^\alpha_\gamma\ \mathcal U^{+1}_F\otimes \mathfrak Q^\gamma_b+\tfrac 12\mathfrak Q^\gamma_b\otimes\mathfrak L^\alpha_\gamma\nonumber\\
&\ -\tfrac 12 \mathfrak R^c_b\ \mathcal U^{+1}_F\otimes \mathfrak Q^\alpha_c+\tfrac 12\mathfrak Q^\alpha_c\otimes\mathfrak R^c_b-\tfrac 12\mathfrak C\ \mathcal U^{+1}_F\otimes\mathfrak Q^\alpha_b+\tfrac 12\mathfrak Q^\alpha_b\otimes\mathfrak C\nonumber\\
&\ +\tfrac 12\varepsilon^{\alpha\gamma}\varepsilon_{bd}\ \mathfrak P\ \mathcal U^{-1}_F\otimes\mathfrak S^d_\gamma-\tfrac 12\varepsilon^{\alpha\gamma}\varepsilon_{bd}\ \mathfrak S^d_\gamma\ \mathcal U^{+2}\otimes\mathfrak P,\nonumber\\
\Delta\widehat{\mathfrak S}^a_\beta=&\widehat{\mathfrak S}^a_\beta\otimes 1+\mathcal U^{-1}_F\otimes \widehat{\mathfrak S}^a_\beta+\tfrac 12 \mathfrak R^a_c\ \mathcal U^{-1}_F\otimes \mathfrak S^c_\beta-\tfrac 12\mathfrak S^c_\beta\otimes\mathfrak R^a_c\nonumber\\
&\ +\tfrac 12 \mathfrak L^\gamma_\beta\ \mathcal U^{-1}_F\otimes \mathfrak S^a_\gamma-\tfrac 12\mathfrak S^a_\gamma\otimes\mathfrak L^\gamma_\beta+\tfrac 12\mathfrak C\ \mathcal U^{-1}_F\otimes\mathfrak S^a_\beta-\tfrac 12\mathfrak S^a_\beta\otimes\mathfrak C\nonumber\\
&\ -\tfrac 12\varepsilon^{ac}\varepsilon_{\beta\delta}\ \mathfrak K\ \mathcal U^{+1}_F\otimes\mathfrak Q^\delta_c+\tfrac 12\varepsilon^{ac}\varepsilon_{\beta\delta}\ \mathfrak Q^\delta_c\ \mathcal U^{-2}\otimes\mathfrak K.\label{coproductsu222}
\end{align}
The action on longer states can be obtained from \eqref{Y222l0} by making use of \eqref{msiteevrep} and incorporating all fermionic grading operators $(-1)^F$ and the braiding factors $\mathcal U$.
The representation of the Yangian $Y[\mathfrak{su}(2|2)]$ without central extension can be obtained by setting the eigenvalues of $\mathfrak P$ and $\mathfrak K$ to 0 and that of $\mathcal U$ to 1.


\paragraph{Two-Particle Undynamic S-Matrix.}

We begin the analysis of the constraints by looking at the two-particle undynamic case, i.e.\ we look for the S-matrix that maps as 
\begin{align}
\text S_{12}(u_{1,2};v_{1,2}):(2|2)_{v_1}\otimes(2|2)_{v_2}\rightarrow(2|2)_{u_1}\otimes (2|2)_{u_2},
\end{align}
with different sets of incoming and outgoing rapidities satisfying $u_1>u_2$ and $v_1<v_2$. 

The level-0 constraints restrict this operator to be of the form 
\begin{align}
\text S_{12}\ket{\phi^a,v_1;\phi^b,v_2}&=A_{12}\ket{\phi^{\{a},u_1;\phi^{b\}},u_2}+B_{12}\ket{\phi^{[a},u_1;\phi^{b]},u_2}+\tfrac 12C_{12}\varepsilon^{ab}\varepsilon_{\alpha\beta}\ket{\psi^\alpha,u_1;\psi_1^\beta,u_2},\nonumber\\
\text S_{12}\ket{\phi^a,v_1;\psi^\beta,v_2}&=D_{12}\ket{\psi^\beta,u_1;\phi^a,u_2}+E_{12}\ket{\phi^a,u_1;\psi^\beta,u_2},
\label{ansatzsu222}
\\
\text S_{12}\ket{\psi^\alpha,v_1;\phi^b,v_2}&=F_{12}\ket{\psi^\alpha,u_1;\phi^b,u_2}+G_{12}\ket{\phi^b,u_1;\psi^\alpha,u_2},\nonumber\\
\text S_{12}\ket{\psi^\alpha,v_1;\psi^\beta,v_2}
&=H_{12}\ket{\psi^{\{\alpha},u_1;\psi^{\beta\}},u_2}+K_{12}\ket{\psi^{[\alpha},u_1;\psi^{\beta]},u_2}+\tfrac 12L_{12}\varepsilon^{\alpha\beta}\varepsilon_{ab}\ket{\phi^a,u_1;\phi^b,u_2},\nonumber
\end{align}
with ten coefficients $A_{12},...,L_{12}$. This ansatz generalizes the ansatz in Table 1 of  \cite{Beisert:2005tm}, where the rapidities are conserved $v_1=u_2$ and $v_2=u_1$.
The coefficients have to satisfy
\begin{align}
&B_{12}=H_{12},& &C_{12}=0,& &D_{12}=\tfrac 12(A_{12}-H_{12}),& &E_{12}=\tfrac 12(A_{12}+H_{12}),&\nonumber\\
&K_{12}=A_{12},& &L_{12}=0,& &G_{12}=\tfrac 12(A_{12}-H_{12}),& &F_{12}=\tfrac 12(A_{12}+H_{12}),\label{AH}
\end{align}
and thus we conclude that the Yangian level-0 constraints fix the S-matrix up to two degrees of freedom corresponding to two Casimirs of $\mathfrak{su}(2|2)$. Note that the construction of the Casimirs from a Killing form $\kappa^{ab}$ of the algebra is non-trivial since the latter vanishes, see \cite{deLeeuw:2010nd}.

The level-1 constraints further restrict the two-particle S-matrix via
\begin{align}
&v_1=u_2,\quad v_2=u_1,& &H_{12}=A_{12}\frac{i-g u_{12}}{i+g u_{12}}.
\label{AH2}
\end{align}
Thus the two-particle S-matrix simply permutes the Hilbert spaces of the particles as
\begin{align}
&\text S_{12}(u_{1,2}): (2|2)_{u_2}\otimes(2|2)_{u_1}\rightarrow (2|2)_{u_1}\otimes(2|2)_{u_2}.
\end{align}
This two-particle S-matrix satisfies the quantum Yang--Baxter equation \eqref{qYBE} which we checked by explicit calculation.


\paragraph{Three-Particle Undynamic S-Matrix.}
We proceed with the three-particle S-matrix S$_{123}$ which denotes the map
\begin{align}
\text S_{123}(u_{1,2,3};v_{1,2,3}):(2|2)_{v_1}\otimes(2|2)_{v_2}\otimes(2|2)_{v_3}\rightarrow(2|2)_{u_1}\otimes (2|2)_{u_2}\otimes(2|2)_{u_3},
\end{align}
with $u_1>u_2>u_3$ and $v_1<v_2<v_3$.

The level-0 constraints from $\mathfrak R^a_b$ and $\mathfrak L^\alpha_\beta$ restrict the S-matrix to be a linear combination of 70 operators similar to \eqref{ansatzsu222}. This ansatz gets further constrained by the remaining level-0 constraints such that ten degrees of freedom remain.
By imposing the level-1 constraints of the Yangian corresponding to the Lie superalgebra $\mathfrak{su}(2|2)$, we further restrict the form of the coefficients and the outgoing rapidities $v_{1,2,3}$ to
\begin{align}
&v_1=u_3,& &v_2=u_2,& &v_3=u_1.
\end{align}
Four of the ten free coefficients in S$_{123}$ have to vanish and another five are related to the remaining single degree of freedom. The resulting S-matrix is fixed up to an overall factor and permutes the particles in a scattering event. By comparing the unique three-particle solution for the S-matrix with the product of three two-particle S-matrices as in \eqref{3fact}, we find that it factorizes. 


\paragraph{Two-Particle Dynamic S-Matrix.}

We now move on with the analysis of the dynamic constraints on S-matrices. The $Y[\mathfrak{su}(2|2)\ltimes\mathbb R^2]$-invariant two-particle S-matrix in its dynamic representation is the map
\begin{align}
&\text S_{12}(u_{1,2},v_{1,2}):(2|2)_{v_1,C(v_1),P(v_1),K(v_1)}\otimes(2|2)_{v_2,C(v_2),P(v_2),K(v_2)}\nonumber\\
&\hspace{4cm}\rightarrow(2|2)_{u_1,C(u_1),P(u_1),K(u_1)}\otimes(2|2)_{u_2,C(u_2),P(u_2),K(u_2)},
\end{align}
with $u_1>u_2$ and $v_1<v_2$.

The level-0 constraints \eqref{level0} for the central charges $\mathfrak C$, $\mathfrak P$ and $\mathfrak K$ impose the following three relations including their eigenvalues $C$, $P$ and $K$:
\begin{align}
 C(u_1)+C(u_2)&=C(v_1)+C(v_2),\nonumber\\
 P(u_1)+P(u_2)U(u_1)^2&=P(v_1)+P(v_2)U(v_1)^2,\nonumber\\
 K(u_1)+\frac{K(u_2)}{U(u_1)^2}&=K(v_1)+\frac{K(v_2)}{U(v_1)^2}.\label{CPK2}
\end{align}
They constrain the outgoing rapidities $v_1$ and $v_2$ and  the coefficients $a,b,c,d$ in \eqref{su22repQS} are conveniently reparametrized via
\begin{align}
&a=\sqrt{g}\gamma,& &b=\sqrt{g}\frac{\alpha}{\gamma}\left(1-\frac{x^+}{x^-}\right),& &c=\sqrt{g}\frac{i\gamma}{\alpha x^+},& &d=\sqrt{g}\frac{x^+}{i\gamma}\left(1-\frac{x^-}{x^+}\right),\label{repara}
\end{align}
with $x^\pm=x^\pm(u), \gamma=\gamma(u)$ and $x^\pm(u)=x(u\pm\frac{i}{2})$.
The parameter $\gamma$ is associated to a relative rescaling between fermions $\psi^\alpha$ and bosons $\phi^a$, while the constant $\alpha$ corresponds to a rescaling of the vacuum field $Z$.  The condition \eqref{adbc} translates into 
\begin{align}
x^++\frac{1}{x^+}-x^--\frac{1}{x^-}=\frac ig,
\label{eq:coupling}
\end{align}
with the model's coupling constant $g$,
and the eigenvalues of the central charges using \eqref{PK} and \eqref{Cad} are
\begin{align}\label{eq:CPKeigen}
&C=-\frac{1}{2}+igx^--igx^+,
& P=g\alpha\left(1-\frac{x^+}{x^-}\right),
&
& K=\frac{g}{\alpha}\left(1-\frac{x^-}{x^+}\right).
\end{align}
Demanding cocommutativity of the coproduct leads to relations between the braiding factor $\mathcal U$ and the central charges $\mathfrak P$ and $\mathfrak K$, see \cite{Beisert:2007ds}. These are solved if the eigenvalue $U$ of $\mathcal U$ satisfies
\begin{align}
U=\sqrt{\frac{x^+}{x^-}},
\end{align}
such that the momentum $p$ and the parameters $x^\pm$  satisfy 
\begin{align}
\frac{x^+}{x^-}=e^{ip}.
\label{eq:momentum}
\end{align}
In these variables the rapidity can be expressed in terms of the new coordinates $x^\pm$ as
\begin{align}
u=\frac g2(x^++x^-)\Big(1+\frac{1}{x^+x^-}\Big).
\end{align}
It is evident that the constraint equations are solved by
\begin{align}
&(1)\quad v_1=u_1,\ v_2=u_2\nonumber\\
&(2)\quad v_1=u_2,\ v_2=u_1.\label{sols}
\end{align}
The first solution is inconsistent with a non-trivial scattering process for which we naturally impose $u_1>u_2$ and $v_1<v_2$. The second corresponds to a scattering solution with the same sets of rapidities in the incoming and outgoing state. Proceeding with this solution by looking at $[\Delta\mathfrak R^a_b,\text S_{12}]=[\Delta\mathfrak L^\alpha_\beta,\text S_{12}]=0$ shows that in this case the S-matrix has to be of the form given in \eqref{ansatzsu222}, i.e.\ we continue with the same ansatz as in the undynamic case. It has ten degrees of freedom. Unlike in the undynamic case the remaining level 0 constraints fully constrain the two-particle S-matrix up to an overall factor.
We obtain for the coefficients in \eqref{ansatzsu222}
\begin{align}
&A_{12}=\text S_{12}^0\frac{x_1^+-x_2^-}{x_1^--x_2^+},\nonumber\\
&B_{12}=\text S_{12}^0\frac{x_1^+-x_2^-}{x_1^--x_2^+}\left(1-2\frac{1-1/x_1^-x_2^+}{1-1/x_1^+x_2^+}\frac{x_1^--x_2^-}{x_1^+-x_2^-}\right),\nonumber\\
&C_{12}=\text S_{12}^0\frac{2\gamma_1\gamma_2U_1}{\alpha x_2^+x_1^+}\frac{1}{1-1/x_1^+x_2^+}\frac{x_1^--x_2^-}{x_1^--x_2^+},\nonumber\\
&D_{12}=\text S_{12}^0\frac{1}{U_2}\frac{x_1^+-x_2^+}{x_1^--x_2^+},\nonumber\\
&E_{12}=\text S_{12}^0\frac{\gamma_2U_1}{\gamma_1U_2}\frac{x_1^+-x_1^-}{x_1^--x_2^+},\nonumber\\
&F_{12}=\text S_{12}^0\frac{\gamma_1}{\gamma_2}\frac{x_2^+-x_2^-}{x_1^--x_2^+},\nonumber\\
&G_{12}=\text S_{12}^0U_1\frac{x_1^--x_2^-}{x_1^--x_2^+}\nonumber\\
&H_{12}=-\text S_{12}^0\frac{U_1}{U_2},\nonumber\\
&K_{12}=-\text S_{12}^0\frac{U_1}{U_2}\left(1-2\frac{1-1/x_1^+x_2^-}{1-1/x_1^-x_2^-}\frac{x_1^+-x_2^+}{x_1^--x_2^+}\right),\nonumber\\
&L_{12}=-\text S_{12}^0\frac{2\alpha(x_2^+-x_2^-)(x_1^+-x_1^-)}{\gamma_1\gamma_2U_2x_2^-x_1^-}\frac{1}{1-1/x_2^-x_1^-}\frac{x_1^+-x_2^+}{x_1^--x_2^+}.\label{Smatrixsu22}
\end{align}
This is the well-known result for the $\mathfrak{su}(2|2)\ltimes\mathbb R^2$-invariant two-particle S-matrix calculated in \cite{Beisert:2005tm}. 

Since the level-0 constraints completely determine this two-particle S-matrix there only remains the question whether this matrix is also Yangian-invariant. Checking the level-1 constraints explicitly using the coproduct structure in \eqref{coproductsu222} shows that this is indeed the case. This was first shown in \cite{Beisert:2007ds}. By numerical analysis, i.e. insertion of different real and complex values for $x^\pm$ and $g$, we furthermore checked that the two-particle S-matrix satisfies the qYBE given in \eqref{qYBE}.

For a conserved set of rapidities, the S-matrix in \eqref{Smatrixsu22} is the unique solution to the Yangian constraints. Analyzing the level-0 and -1 constraint equations for different sets of incoming and outgoing rapidities, e.g.\ the equations \eqref{CPK2}, to check whether this is the only solution, becomes untractable in \texttt{Mathematica}. 
In the two-particle case we can circumvent this problem by using the explicit AdS/CFT form of the above rapidity dependent functions. In the context of $\superN=4$ SYM theory the solutions $x^\pm=x^\pm(u)$ of \eqref{eq:coupling} take the form \cite{Beisert:2004hm}%
\begin{align}
x(u)&=\frac{u}{2g}+\frac{u}{2g}\sqrt{1-\frac{4g^2}{u^2}},
&
u(x)&=g\Big(x+\frac{1}{x}\Big).
\label{solxu}
\end{align}
It is most convenient to express the functions in terms of momenta in order to straightforwardly implement the constraint of momentum conservation. We have (cf.\ \eqref{eq:Calphabeta})
\begin{align}
u(p)&=\frac{1}{2} \cot \frac{p}{2} \sqrt{1+16 g^2 \sin^2\frac{p}{2}},
&
C(p)&= \frac{1}{2}
\sqrt{1+16 g^2 \sin^2 \frac{p}{2}}.
\label{eq:uCAdSCFT}
\end{align}
Then we numerically verified that the constraints of momentum and central charge (energy) conservation, namely \eqref{eq:pEconservation}, imply conservation of individual momenta (up to discrete shifts). Here the momenta $p_{1,2}$ and $q_{1,2}$ are associated with the rapidities $v_{1,2}$ and $u_{1,2}$, respectively, and we have specified random numerical values for $p_{1,2}$ and $g$ and solved for $q_{1,2}$. 
Thus the Yangian constraints corresponding to $\mathfrak{su}(2|2)\ltimes\mathbb R^2$ together with \eqref{eq:uCAdSCFT} imply the conservation of individual momenta/rapidities in two-particle scattering events. 


\paragraph{Three-Particle Dynamic S-Matrix.}

The $Y[\mathfrak{su}(2|2)\ltimes\mathbb R^2]$-invariant three-particle S-matrix in its dynamic representation is the map
\begin{align}
&\text S_{123}(u_{i},v_{i}):(2|2)_{v_1,C(v_1),P(v_1),K(v_1)}\otimes(2|2)_{v_2,C(v_2),P(v_2),K(v_2)}\otimes(2|2)_{v_3,C(v_3),P(v_3),K(v_3)}\nonumber\\
&\hspace{1.8cm}\rightarrow(2|2)_{u_1,C(u_1),P(u_1),K(u_1)}\otimes(2|2)_{u_2,C(u_2),P(u_2),K(u_2)}\otimes(2|2)_{u_3,C(u_3),P(u_3),K(u_3)}
\end{align}
with $u_1>u_2>u_3$ and $v_1<v_2<v_3$.

Similar to the two-particle case, the Yangian constraints imply relations among functions of the incoming and outgoing rapidities. However, due to the complicated constraints we were not able to verify whether the constraint equations imply the conservation of the set of rapidities. 
Obviously, they are solved if the set of outgoing rapidities equals the set of incoming rapidities. The case corresponding to the true three-particle scattering with $u_1>u_2>u_3$ and $v_1<v_2<v_3$ is
\begin{align}\label{eq:threepartcons}
v_1=u_3,\quad v_2=u_2,\quad v_3=u_1.
\end{align}
Demanding vanishing commutators $[\Delta^2\mathfrak R^a_b,\text S_{123}]$ and $[\Delta^2\mathfrak L^\alpha_\beta,\text S_{123}]$ restricts the three-particle S-matrix S$_{123}$ to be of the same form as in the discussion of the undynamic Yangian constraints in the previous section with 70 free coefficients. The remaining level-0 constraints reduce this number to 2. We did not do these calculations analytically due to the great amount of computational power needed, but evaluated the relations numerically by setting the $x_i^\pm$ and $g$ to arbitrary values respecting \eqref{eq:coupling}. Note that this number of degrees of freedom corresponds to the expectations from the discussion of the representation theory of this algebra, see \cite{Beisert:2006qh,Puletti:2007hq}. There it is shown that the tensor product of three one-particle states denoted by $\braket{m=0,n=0,\vec C}$ with $\mathfrak{su}(2)\times\mathfrak{su}(2)$ Dynkin labels $m$ and $n$ and  eigenvalues $\vec C=(C,P,K)$ of the central charges decomposes as
\begin{align}
\braket{0,0,\vec C_1}\otimes\braket{0,0,\vec C_2}\otimes\braket{0,0,\vec C_3}=\{1,0,\vec C_1+\vec C_2+\vec C_3\}\oplus\{0,1,\vec C_1+\vec C_2+\vec C_3\}.
\end{align}
The bracket $\braket{..}$ denotes a state from a long multiplet and $\{..\}$ one from a short multiplet with $C^2-PK=\tfrac 14(n+m+1)^2$. Since we found the three-particle $\mathfrak{su}(2|2)\ltimes\mathbb{R}^2$-invariant S-matrix numerically, we proceed with a numerical analysis of the level-1 constraints of $Y[\mathfrak{su}(2|2)\ltimes\mathbb R^2]$. They are only fulfilled if the two remaining free coefficients are related to each other. This relation is the same for all the commutators such that we are left with a three-particle S-matrix that is determined up to an overall factor. We compared our numerical result with a product of three two-particle S-matrices and the equality of both expressions confirms that the three-particle S-matrix factorizes. 

Note that the two-particle dynamic S-matrix reduces to the two-particle undynamic S-matrix in the limit where all eigenvalues $U_i$ of $\mathcal U$ and the $\gamma_i$ are set to $1$ and the remaining ingredients in \eqref{Smatrixsu22} are expanded in powers of the coupling constant $g$ (with $\alpha=\mathcal O(g)$). Notice that the number of degrees of freedom after exploiting the level-0 Yangian constraints differs both in the two- and three-particle case with $Y[\mathfrak{su}(2|2)\ltimes\mathbb R^2]$ in the dynamic representation being more restrictive than $Y[\mathfrak{su}(2|2)]$ in the undynamic representation.


\paragraph{$\mathfrak{u}(1|1)$ from $\mathfrak{su}(2|2)$.}

Note that we may obtain results derived in the previous section on the $Y[\mathfrak{u}(1|1)]$-invariant S-matrices from the discussion of $Y[\mathfrak{su}(2|2)(\ltimes\mathbb R^2)]$. Indeed, the fundamental representation of $\mathfrak{u}(1|1)$ in \eqref{su11fb} can be obtained from the $\mathfrak{su}(2|2)$ representation \eqref{su22repRL}-\eqref{adbc} by restricting the action of the $\mathfrak{su}(2|2)$-generators to a single boson and a single fermion, i.e.\ we set all indices to $a=b=\alpha=\beta=1$ and identify $\ket{\phi}:=\ket{\phi^1}$ and $\ket{\psi}:=\ket{\psi^1}$. Then the $\mathfrak{u}(1|1)$ generators $\mathfrak Q$ and $\mathfrak S$ can be obtained from  $\mathfrak Q^1_1$ and $\mathfrak S^1_1$ by relating the representation coefficients $a$ and $d$ in \eqref{su22repQS} to $\mathfrak{u}(1|1)$-coefficients via $a=q$, and $d=\frac cq$. Note that the $\mathfrak{su}(2|2)$ coefficients are constrained via \eqref{adbc}. For the undynamic case this implies together with $P=K=0$ that we only obtain the case where the $\mathfrak{u}(1|1)$ central charge eigenvalue is set to $c=1$. The outer automorphism $\mathfrak B$ of $\mathfrak{su}(1|1)$ can be obtained from the linear combination 
\begin{align}
2(b-1)\mathfrak{L}^1_1+2(b+1)\mathfrak{R}^1_1
\end{align}
where $b$ is the $\mathfrak{u}(1|1)$ hypercharge. The central charge $\mathfrak C$ of $\mathfrak{u}(1|1)$ is given by the combination of $\mathfrak{su}(2|2)$ generators of the form
\begin{align}
\mathfrak C+\mathfrak R^1_1+\mathfrak L^1_1.
\end{align}
These formulae can also be used to derive the eigenvalues of $\mathfrak{B}$ and $\mathfrak{C}$ of $\mathfrak{u}(1|1)$ from the eigenvalues of the $\mathfrak{su}(2|2)$  generators.
This close connection of both algebras allows us to derive the $Y[\mathfrak{u}(1|1)]$-invariant S-matrices discussed in the previous section from the $\mathfrak{su}(2|2)$ results. We checked that both the undynamic and the dynamic two-particle S-matrices are compatible, which implies via consistent factorization that also the three-particle S-matrices agree in this limit.


Let us summarize the results of this subsection. Using the undynamic Yangian constraints on S-matrices we found that the sets of rapidities in a two- and three-particle scattering event are conserved. Also in the dynamic setup the conservation of the sets of rapidities solves the Yangian constraints. Their uniqueness for two-particle scattering can be deduced from momentum and central charge (energy) conservation in the context of {\sym} theory. In the case of three-particle scattering we were not able to verify that Yangian symmetry implies the conservation of the set of rapidities due to the complicated constraints that could not be solved using computer algebra.
For conserved sets of rapidities, we determined both the two-particle undynamic $Y[\mathfrak{su}(2|2)]$- and dynamic $Y[\mathfrak{su}(2|2)\ltimes\mathbb R^2]$-invariant S-matrix \eqref{ansatzsu222} with coefficients \eqref{AH,AH2} and \eqref{Smatrixsu22}, respectively, up to the overall dressing factor. In the undynamic case we also determined the three-particle S-matrix and checked that it factorizes consistently into three two-particle S-matrices. In the dynamic case we checked the qYBE as well as the factorization of the three-particle S-matrix numerically. While the resulting undynamic S-matrices are of difference form, the dynamic ones are not, since the Yangian coproduct does not have the standard structure used in the arguments above \eqref{eq:diffform}.


\section{Conclusions and Future Directions}

In this paper we have investigated the implications of Yangian symmetry on the three-particle S-matrix in two dimensions. To be explicit, we studied the three Yangian algebras $Y[\mathfrak{su}(N)]$, $Y[\mathfrak{u}(1|1)]$ and $Y[\mathfrak{su}(2|2)\ltimes\mathbb R^2]$ and found that the resulting Yangian constraints imply factorization. In doing so, we partially employed numerical values for some of the parameters in order to solve the involved constraint equations depending on arbitrary coefficients, or functions of the rapidities, cf.\ \tabref{table:results}. 
While in some cases we were able to show that conservation of the set of rapidities follows from Yangian symmetry, for the cases of the dynamic  three-particle scattering we added this point as an assumption in order to solve the intricate constraint equations.

Since Yangian symmetry and factorized scattering are just two different formulations of quantum integrability, their relation observed here is not unexpected. However, due to the lack of a generic definition of integrability  it is important to clarify this fundamental question, in particular in the context of the less explored dynamic representations motivated by the AdS/CFT duality. The tremendous recent progress on integrable bootstrap approaches towards AdS/CFT observables (see e.g.\ \cite{Basso:2013vsa,Basso:2015zoa,Eden:2016xvg,Fleury:2016ykk}) underlines the need for better understanding the Yangian as their algebraic foundation.

Let us finish by pointing out some particularly interesting future directions. 
An important point would be to extend the analysis of this paper to noncompact representations. A direct application is to understand in which sense the Yangian symmetry for higher dimensional S-matrices implies factorization, e.g.\ for the Yangian symmetry of amplitudes in $\superN=4$ SYM theory \cite{Drummond:2009fd,Bargheer:2009qu}, in ABJM theory \cite{Bargheer:2010hn} or in the recently found fishnet theories \cite{Chicherin:2017cns,Chicherin:2017frs}. This should help to clarify the question mark in \tabref{tab:symresults} in the introduction.

As demonstrated above, conventional Yangian symmetry implies that the two-particle S-matrix only depends on the difference of rapidities. Hence, the conventional Yangian is tied to an underlying boost symmetry of the physical model, cf.\ e.g.\ \cite{Loebbert:2016cdm}. On the other hand, the dynamic representations of AdS/CFT include non-trivial braiding factors which modify the standard Yangian constraints and imply that the S-matrix is not of difference form. It would be very interesting to understand whether there is an AdS/CFT generalization of the two-dimensional Lorentz boost that shifts the spectral parameter of the model in analogy to the boost automorphism of the Yangian and what it tells us about the AdS/CFT quantum group. Recently, there has been very interesting progress into this direction that should further be explored \cite{Borsato:2017lpf,Beisert:2017xqx} (see also the earlier works \cite{Gomez:2007zr,Young:2007wd}).

An important reformulation of quantum integrability in the language of gauge theories was recently proposed by Costello, Yamazaki and Witten \cite{Costello:2013zra,Costello:2017dso,Costello:2018gyb}. This includes the conventional Yangian algebra representing rational solutions to the quantum Yang--Baxter equation. It would be very interesting to also understand how the dynamic representations considered in this paper embed into their formalism and to think about factorization from their point of view.

Finally, it would be important to provide a general proof that Yangian symmetry, in conventional or dynamic representations, implies factorization of the S-matrix. While desirable to mimic the beautiful intuitive arguments that are underlying the proof for local higher charges, it seems that nonlocal charges require more technical reasoning.


\paragraph{Acknowledgments.}

We are grateful to
Niklas Beisert,
Ben Hoare,
Thomas Klose,
Marius de Leeuw,
Marc Magro,
Jean-Michel Maillet,
Tristan McLoughlin,
Matteo Rosso,
Pedro Vieira and
Masahito Yamazaki for very interesting and insightful discussions on the topics of this paper. We thank Marius de Leeuw for helpful comments on the manuscript as well as Tristan McLoughlin for important remarks on the first version of the preprint.
The work of FL is funded by the Deutsche Forschungsgemeinschaft (DFG, German Research Foundation) -- Projektnummer 363895012. The work of AS is supported by the SFI grant 15/CDA/3472.

\pdfbookmark[1]{\refname}{references}
\bibliographystyle{nb}
\bibliography{YangFacto}

\begin{thebibliography}{10}
\providecommand{\href}[2]{#2}
\providecommand{\arxivref}[2]{\href{http://arxiv.org/abs/#1}{#2}}
\providecommand{\doiref}[2]{\href{http://dx.doi.org/#1}{#2}}
\providecommand{\nbbstauthor}[1]{#1}
\providecommand{\nbbstjournal}[1]{\textsf{#1}}
\providecommand{\nbbsttitle}[1]{\textit{#1}}
\providecommand{\nbbsturl}[1]{\texttt{#1}}
\providecommand{\nbbsteprint}[1]{\texttt{#1}}
\providecommand{\nbbststyle}{\raggedright\small\parskip0pt}
\nbbststyle

\bibitem{Kulish:1975ba}
\nbbstauthor{P.~P.~Kulish},
\nbbsttitle{``{Factorization of the Classical and Quantum s Matrix and
  Conservation Laws}''},
\nbbstjournal{\doiref{10.1007/BF01079418}{Theor.~Math.~Phys.~26,~132~(1976)}},
[Teor. Mat. Fiz.26,198(1976)].

\bibitem{Shankar:1977cm}
\nbbstauthor{R.~Shankar and E.~Witten},
\nbbsttitle{``{The S-Matrix of the Supersymmetric Nonlinear Sigma Model}''},
\nbbstjournal{\doiref{10.1103/PhysRevD.17.2134}{Phys.~Rev.~D17,~2134~(1978)}}.

\bibitem{Iagolnitzer:1977sw}
\nbbstauthor{D.~Iagolnitzer},
\nbbsttitle{``{Factorization of the Multiparticle s Matrix in Two-Dimensional
  Space-Time Models}''},
\nbbstjournal{\doiref{10.1103/PhysRevD.18.1275}{Phys.~Rev.~D18,~1275~(1978)}}.

\bibitem{Iagolnitzer:1978my}
\nbbstauthor{D.~Iagolnitzer},
\nbbsttitle{``{The Multiparticle S Matrix in Two-Dimensional Space-Time
  Models}''},
\nbbstjournal{\doiref{10.1016/0370-2693(78)90277-0}{Phys.~Lett.~76B,~207~(1978)}}.

\bibitem{Parke:1980ki}
\nbbstauthor{S.~J.~Parke},
\nbbsttitle{``{Absence of Particle Production and Factorization of the $S$
  Matrix in (1+1)-dimensional Models}''},
\nbbstjournal{\doiref{10.1016/0550-3213(80)90196-0}{Nucl.~Phys.~B174,~166~(1980)}}.

\bibitem{Bernard:1992ya}
\nbbstauthor{D.~Bernard},
\nbbsttitle{``{An Introduction to Yangian Symmetries}''},
\nbbstjournal{\doiref{10.1142/S0217979293003371}{Int.~J.~Mod.~Phys.~B7,~3517~(1993)}},
\nbbsteprint{\arxivref{hep-th/9211133}{hep-th/9211133}},
in: \nbbsttitle{``{Advanced Research Workshop on Integrable Quantum Field
  Theories: Conformal Field Theories and Current Algebra, Integrable Models, 2D
  Quantum Gravity, Matrix Models and String Theory Como, Italy, September
  14-19, 1992}''},
pp.~3517-3530.

\bibitem{MacKay:2004tc}
\nbbstauthor{N.~J.~MacKay},
\nbbsttitle{``{Introduction to Yangian symmetry in integrable field theory}''},
\nbbstjournal{\doiref{10.1142/S0217751X05022317}{Int.~J.~Mod.~Phys.~A20,~7189~(2005)}},
\nbbsteprint{\arxivref{hep-th/0409183}{hep-th/0409183}},
in: \nbbsttitle{``{ESI Workshop on String Theory on Non-Compact and
  Time-Dependent Backgrounds Vienna, Austria, June 7-18, 2004}''},
pp.~7189-7218.

\bibitem{Torrielli:2011gg}
\nbbstauthor{A.~Torrielli},
\nbbsttitle{``{Yangians, S-matrices and AdS/CFT}''},
\nbbstjournal{\doiref{10.1088/1751-8113/44/26/263001}{J.~Phys.~A44,~263001~(2011)}},
\nbbsteprint{\arxivref{1104.2474}{arxiv:1104.2474}}.

\bibitem{Loebbert:2016cdm}
\nbbstauthor{F.~Loebbert},
\nbbsttitle{``{Lectures on Yangian Symmetry}''},
\nbbstjournal{\doiref{10.1088/1751-8113/49/32/323002}{J.~Phys.~A49,~323002~(2016)}},
\nbbsteprint{\arxivref{1606.02947}{arxiv:1606.02947}}.

\bibitem{MacKay:1998kx}
\nbbstauthor{N.~J.~MacKay},
\nbbsttitle{``{Local versus nonlocal charges in integrable field theories}''},
\nbbstjournal{\doiref{10.1023/A:1021621709683}{Czech.~J.~Phys.~48,~1441~(1998)}},
in: \nbbsttitle{``{Quantum groups and integrable systems. Proceedings, 7th
  International Colloquium on Quantum Groups, Prague, Czech Republic, June
  18-20, 1998}''},
pp.~1441-1446.

\bibitem{Tarasov:1983cj}
\nbbstauthor{V.~O.~Tarasov, L.~A.~Takhtajan and L.~D.~Faddeev},
\nbbsttitle{``{Local Hamiltonians for integrable quantum models on a
  lattice}''},
\nbbstjournal{\doiref{10.1007/BF01018648}{Theor.~Math.~Phys.~57,~1059~(1983)}},
[Teor. Mat. Fiz.57,163(1983)].

\bibitem{Drummond:2009fd}
\nbbstauthor{J.~M.~Drummond, J.~M.~Henn and J.~Plefka},
\nbbsttitle{``{Yangian symmetry of scattering amplitudes in N=4 super
  Yang-Mills theory}''},
\nbbstjournal{\doiref{10.1088/1126-6708/2009/05/046}{JHEP~0905,~046~(2009)}},
\nbbsteprint{\arxivref{0902.2987}{arxiv:0902.2987}},
in: \nbbsttitle{``{Strangeness in quark matter. Proceedings, International
  Conference, SQM 2008, Beijing, P.R. China, October 5-10, 2008}''},
pp.~046.

\bibitem{Chicherin:2017cns}
\nbbstauthor{D.~Chicherin, V.~Kazakov, F.~Loebbert, D.~Muller and D.-l.~Zhong},
\nbbsttitle{``{Yangian Symmetry for Bi-Scalar Loop Amplitudes}''},
\nbbstjournal{\doiref{10.1007/JHEP05(2018)003}{JHEP~1805,~003~(2018)}},
\nbbsteprint{\arxivref{1704.01967}{arxiv:1704.01967}}.

\bibitem{Chicherin:2017frs}
\nbbstauthor{D.~Chicherin, V.~Kazakov, F.~Loebbert, D.~Muller and D.-l.~Zhong},
\nbbsttitle{``{Yangian Symmetry for Fishnet Feynman Graphs}''},
\nbbstjournal{\doiref{10.1103/PhysRevD.96.121901}{Phys.~Rev.~D96,~121901~(2017)}},
\nbbsteprint{\arxivref{1708.00007}{arxiv:1708.00007}}.

\bibitem{Bargheer:2010hn}
\nbbstauthor{T.~Bargheer, F.~Loebbert and C.~Meneghelli},
\nbbsttitle{``{Symmetries of Tree-level Scattering Amplitudes in N=6
  Superconformal Chern-Simons Theory}''},
\nbbstjournal{\doiref{10.1103/PhysRevD.82.045016}{Phys.~Rev.~D82,~045016~(2010)}},
\nbbsteprint{\arxivref{1003.6120}{arxiv:1003.6120}}.

\bibitem{Drinfeld:1985rx}
\nbbstauthor{V.~Drinfeld},
\nbbsttitle{``{Hopf algebras and the quantum Yang-Baxter equation}''},
\nbbstjournal{Sov.~Math.~Dokl.~32,~254~(1985)}.

\bibitem{Dorey:1996gd}
\nbbstauthor{P.~Dorey},
\nbbsttitle{``{Exact S matrices}''},
\nbbsteprint{\arxivref{hep-th/9810026}{hep-th/9810026}},
in: \nbbsttitle{``{Conformal field theories and integrable models. Proceedings,
  Eotvos Graduate Course, Budapest, Hungary, August 13-18, 1996}''},
pp.~85-125.

\bibitem{Luscher:1977uq}
\nbbstauthor{M.~Luscher},
\nbbsttitle{``{Quantum Nonlocal Charges and Absence of Particle Production in
  the Two-Dimensional Nonlinear Sigma Model}''},
\nbbstjournal{\doiref{10.1016/0550-3213(78)90211-0}{Nucl.~Phys.~B135,~1~(1978)}}.

\bibitem{Woo:1979nj}
\nbbstauthor{C.~H.~Woo},
\nbbsttitle{``{Nonlocal Charges in Two Dimensions}''},
\nbbstjournal{\doiref{10.1103/PhysRevD.20.1880}{Phys.~Rev.~D20,~1880~(1979)}}.

\bibitem{Curtright:1994be}
\nbbstauthor{T.~Curtright and C.~K.~Zachos},
\nbbsttitle{``{Currents, charges, and canonical structure of pseudodual chiral
  models}''},
\nbbstjournal{\doiref{10.1103/PhysRevD.49.5408}{Phys.~Rev.~D49,~5408~(1994)}},
\nbbsteprint{\arxivref{hep-th/9401006}{hep-th/9401006}}.

\bibitem{Nappi:1979ig}
\nbbstauthor{C.~R.~Nappi},
\nbbsttitle{``{Some Properties of an Analog of the Nonlinear $\sigma$
  Model}''},
\nbbstjournal{\doiref{10.1103/PhysRevD.21.418}{Phys.~Rev.~D21,~418~(1980)}}.

\bibitem{Arutyunov:2009ga}
\nbbstauthor{G.~Arutyunov and S.~Frolov},
\nbbsttitle{``{Foundations of the $\text{AdS}_5 \times \text{S}^5$ Superstring.
  Part I}''},
\nbbstjournal{\doiref{10.1088/1751-8113/42/25/254003}{J.~Phys.~A42,~254003~(2009)}},
\nbbsteprint{\arxivref{0901.4937}{arxiv:0901.4937}}.

\bibitem{Beisert:2010jr}
\nbbstauthor{N.~Beisert et~al.},
\nbbsttitle{``{Review of AdS/CFT Integrability: An Overview}''},
\nbbstjournal{\doiref{10.1007/s11005-011-0529-2}{Lett.~Math.~Phys.~99,~3~(2012)}},
\nbbsteprint{\arxivref{1012.3982}{arxiv:1012.3982}}.

\bibitem{Bombardelli:2016rwb}
\nbbstauthor{D.~Bombardelli, A.~Cagnazzo, R.~Frassek, F.~Levkovich-Maslyuk,
  F.~Loebbert, S.~Negro, I.~M.~Szecsenyi, A.~Sfondrini, S.~J.~van~Tongeren and
  A.~Torrielli},
\nbbsttitle{``{An integrability primer for the gauge-gravity correspondence: An
  introduction}''},
\nbbstjournal{\doiref{10.1088/1751-8113/49/32/320301}{J.~Phys.~A49,~320301~(2016)}},
\nbbsteprint{\arxivref{1606.02945}{arxiv:1606.02945}}.

\bibitem{Beisert:2005tm}
\nbbstauthor{N.~Beisert},
\nbbsttitle{``{The SU(2|2) dynamic S-matrix}''},
\nbbstjournal{\doiref{10.4310/ATMP.2008.v12.n5.a1}{Adv.~Theor.~Math.~Phys.~12,~945~(2008)}},
\nbbsteprint{\arxivref{hep-th/0511082}{hep-th/0511082}}.

\bibitem{Shastry:1986zz}
\nbbstauthor{B.~S.~Shastry},
\nbbsttitle{``{Exact Integrability of the One-Dimensional Hubbard Model}''},
\nbbstjournal{\doiref{10.1103/PhysRevLett.56.2453}{Phys.~Rev.~Lett.~56,~2453~(1986)}}.

\bibitem{Beisert:2006qh}
\nbbstauthor{N.~Beisert},
\nbbsttitle{``{The Analytic Bethe Ansatz for a Chain with Centrally Extended
  su(2|2) Symmetry}''},
\nbbstjournal{\doiref{10.1088/1742-5468/2007/01/P01017}{J.~Stat.~Mech.~0701,~P01017~(2007)}},
\nbbsteprint{\arxivref{nlin/0610017}{nlin/0610017}}.

\bibitem{Martins:2007hb}
\nbbstauthor{M.~J.~Martins and C.~S.~Melo},
\nbbsttitle{``{The Bethe ansatz approach for factorizable centrally extended
  S-matrices}''},
\nbbstjournal{\doiref{10.1016/j.nuclphysb.2007.05.021}{Nucl.~Phys.~B785,~246~(2007)}},
\nbbsteprint{\arxivref{hep-th/0703086}{hep-th/0703086}}.

\bibitem{Staudacher:2004tk}
\nbbstauthor{M.~Staudacher},
\nbbsttitle{``{The Factorized S-matrix of CFT/AdS}''},
\nbbstjournal{\doiref{10.1088/1126-6708/2005/05/054}{JHEP~0505,~054~(2005)}},
\nbbsteprint{\arxivref{hep-th/0412188}{hep-th/0412188}}.

\bibitem{Puletti:2007hq}
\nbbstauthor{V.~Giangreco Marotta~Puletti, T.~Klose and O.~Ohlsson~Sax},
\nbbsttitle{``{Factorized world-sheet scattering in near-flat AdS(5) x
  S**5}''},
\nbbstjournal{\doiref{10.1016/j.nuclphysb.2007.09.018}{Nucl.~Phys.~B792,~228~(2008)}},
\nbbsteprint{\arxivref{0707.2082}{arxiv:0707.2082}}.

\bibitem{Beisert:2007ds}
\nbbstauthor{N.~Beisert},
\nbbsttitle{``{The S-matrix of AdS / CFT and Yangian symmetry}''},
\nbbstjournal{PoS~SOLVAY,~002~(2006)},
\nbbsteprint{\arxivref{0704.0400}{arxiv:0704.0400}},
in: \nbbsttitle{``{Proceedings, Bethe ansatz: 75 years later: Brussels,
  Belgium, 19-21 October 2006}''},
pp.~002.

\bibitem{Arutyunov:2006yd}
\nbbstauthor{G.~Arutyunov, S.~Frolov and M.~Zamaklar},
\nbbsttitle{``{The Zamolodchikov-Faddeev algebra for AdS(5) x S**5
  superstring}''},
\nbbstjournal{\doiref{10.1088/1126-6708/2007/04/002}{JHEP~0704,~002~(2007)}},
\nbbsteprint{\arxivref{hep-th/0612229}{hep-th/0612229}}.

\bibitem{Arutyunov:2008zt}
\nbbstauthor{G.~Arutyunov and S.~Frolov},
\nbbsttitle{``{The S-matrix of String Bound States}''},
\nbbstjournal{\doiref{10.1016/j.nuclphysb.2008.06.005}{Nucl.~Phys.~B804,~90~(2008)}},
\nbbsteprint{\arxivref{0803.4323}{arxiv:0803.4323}}.

\bibitem{deLeeuw:2008dp}
\nbbstauthor{M.~de~Leeuw},
\nbbsttitle{``{Bound States, Yangian Symmetry and Classical r-matrix for the
  AdS(5) x S**5 Superstring}''},
\nbbstjournal{\doiref{10.1088/1126-6708/2008/06/085}{JHEP~0806,~085~(2008)}},
\nbbsteprint{\arxivref{0804.1047}{arxiv:0804.1047}}.

\bibitem{Komatsu:2017buu}
\nbbstauthor{S.~Komatsu},
\nbbsttitle{``{Lectures on Three-point Functions in N=4 Supersymmetric
  Yang-Mills Theory}''},
\nbbsteprint{\arxivref{1710.03853}{arxiv:1710.03853}}.

\bibitem{Basso:2015zoa}
\nbbstauthor{B.~Basso, S.~Komatsu and P.~Vieira},
\nbbsttitle{``{Structure Constants and Integrable Bootstrap in Planar N=4 SYM
  Theory}''},
\nbbsteprint{\arxivref{1505.06745}{arxiv:1505.06745}}.

\bibitem{Beisert:2005wm}
\nbbstauthor{N.~Beisert},
\nbbsttitle{``{An SU(1|1)-invariant S-matrix with dynamic representations}''},
\nbbstjournal{Bulg.~J.~Phys.~33S1,~371~(2006)},
\nbbsteprint{\arxivref{hep-th/0511013}{hep-th/0511013}},
in: \nbbsttitle{``{4th International Symposium on Quantum Theory and Symmetries
  and 6th International Workshop on Lie Theory and Its Applications in Physics
  (QTS-4) (LT-6) Varna, Bulgaria, August 15-21, 2005}''},
pp.~371-381.

\bibitem{Gomez:2006va}
\nbbstauthor{C.~Gomez and R.~Hernandez},
\nbbsttitle{``{The Magnon kinematics of the AdS/CFT correspondence}''},
\nbbstjournal{\doiref{10.1088/1126-6708/2006/11/021}{JHEP~0611,~021~(2006)}},
\nbbsteprint{\arxivref{hep-th/0608029}{hep-th/0608029}}.

\bibitem{Plefka:2006ze}
\nbbstauthor{J.~Plefka, F.~Spill and A.~Torrielli},
\nbbsttitle{``{On the Hopf algebra structure of the AdS/CFT S-matrix}''},
\nbbstjournal{\doiref{10.1103/PhysRevD.74.066008}{Phys.~Rev.~D74,~066008~(2006)}},
\nbbsteprint{\arxivref{hep-th/0608038}{hep-th/0608038}}.

\bibitem{Beisert:2014hya}
\nbbstauthor{N.~Beisert and M.~de~Leeuw},
\nbbsttitle{``{The RTT realization for the deformed $\mathfrak {gl}(2\vert2)$
  Yangian}''},
\nbbstjournal{\doiref{10.1088/1751-8113/47/30/305201}{J.~Phys.~A47,~305201~(2014)}},
\nbbsteprint{\arxivref{1401.7691}{arxiv:1401.7691}}.

\bibitem{Khoroshkin:1994uk}
\nbbstauthor{S.~M.~Khoroshkin and V.~N.~Tolstoi},
\nbbsttitle{``{Yangian double and rational R matrix}''},
\nbbsteprint{\arxivref{hep-th/9406194}{hep-th/9406194}}.

\bibitem{2005math......4302S}
\nbbstauthor{V.~{Stukopin}},
\nbbsttitle{``{Quantum Double of Yangian of Lie Superalgebra $A(m,n)$ and
  computation of Universal $R$-matrix}''},
\nbbstjournal{ArXiv~Mathematics~e-prints~A47,~V.~{Stukopin}~(2005)},
\nbbsteprint{\arxivref{math/0504302}{math/0504302}}.

\bibitem{Rej:2010mu}
\nbbstauthor{A.~Rej and F.~Spill},
\nbbsttitle{``{The Yangian of sl(n|m) and the universal R-matrix}''},
\nbbstjournal{\doiref{10.1007/JHEP05(2011)012}{JHEP~1105,~012~(2011)}},
\nbbsteprint{\arxivref{1008.0872}{arxiv:1008.0872}}.

\bibitem{Moriyama:2007jt}
\nbbstauthor{S.~Moriyama and A.~Torrielli},
\nbbsttitle{``{A Yangian double for the AdS/CFT classical r-matrix}''},
\nbbstjournal{\doiref{10.1088/1126-6708/2007/06/083}{JHEP~0706,~083~(2007)}},
\nbbsteprint{\arxivref{0706.0884}{arxiv:0706.0884}}.

\bibitem{Beisert:2016qei}
\nbbstauthor{N.~Beisert, M.~de~Leeuw and R.~Hecht},
\nbbsttitle{``{Maximally extended sl(2|2) as a quantum double}''},
\nbbstjournal{\doiref{10.1088/1751-8113/49/43/434005}{J.~Phys.~A49,~434005~(2016)}},
\nbbsteprint{\arxivref{1602.04988}{arxiv:1602.04988}}.

\bibitem{Reshetikhin:1990ep}
\nbbstauthor{N.~Reshetikhin},
\nbbsttitle{``{Multiparameter quantum groups and twisted quasitriangular Hopf
  algebras}''},
\nbbstjournal{\doiref{10.1007/BF00626530}{Lett.~Math.~Phys.~20,~331~(1990)}}.

\bibitem{Spill:2012qe}
\nbbstauthor{F.~Spill},
\nbbsttitle{``{Yangians in Integrable Field Theories, Spin Chains and
  Gauge-String Dualities}''},
\nbbstjournal{\doiref{10.1142/S0129055X12300014}{Rev.~Math.~Phys.~24,~1230001~(2012)}},
\nbbsteprint{\arxivref{1201.1884}{arxiv:1201.1884}},
[Rev. Math. Phys.24,0001(2012)].

\bibitem{Chari:1994pz}
\nbbstauthor{V.~Chari and A.~Pressley},
\nbbsttitle{``{A guide to quantum groups}''}.

\bibitem{Bombardelli:2016scq}
\nbbstauthor{D.~Bombardelli},
\nbbsttitle{``{S-matrices and integrability}''},
\nbbstjournal{\doiref{10.1088/1751-8113/49/32/323003}{J.~Phys.~A49,~323003~(2016)}},
\nbbsteprint{\arxivref{1606.02949}{arxiv:1606.02949}}.

\bibitem{Essler:1992py}
\nbbstauthor{F.~H.~L.~Essler, V.~E.~Korepin and K.~Schoutens},
\nbbsttitle{``{New exactly solvable model of strongly correlated electrons
  motivated by high T(c) superconductivity}''},
\nbbstjournal{\doiref{10.1103/PhysRevLett.68.2960}{Phys.~Rev.~Lett.~68,~2960~(1992)}},
\nbbsteprint{\arxivref{cond-mat/9209002}{cond-mat/9209002}}.

\bibitem{Essler:1992bn}
\nbbstauthor{F.~H.~L.~Essler, V.~E.~Korepin and K.~Schoutens},
\nbbsttitle{``{Electronic model for superconductivity}''},
\nbbstjournal{\doiref{10.1103/PhysRevLett.70.73}{Phys.~Rev.~Lett.~70,~73~(1993)}}.

\bibitem{Essler:1992uc}
\nbbstauthor{F.~H.~L.~Essler, V.~E.~Korepin and K.~Schoutens},
\nbbsttitle{``{Exact solution of an electronic model of superconductivity in
  (1+1)-dimensions. 1.}''},
\nbbstjournal{\doiref{10.1142/S0217979294001354}{Int.~J.~Mod.~Phys.~B8,~3205~(1994)}},
\nbbsteprint{\arxivref{cond-mat/9211001}{cond-mat/9211001}}.

\bibitem{deLeeuw:2010nd}
\nbbstauthor{M.~de~Leeuw},
\nbbsttitle{``{The S-matrix of the $AdS_5 x S^5$ superstring}''},
\nbbsteprint{\arxivref{1007.4931}{arxiv:1007.4931}},
\href{https://inspirehep.net/record/863349/files/arXiv:1007.4931.pdf}{\nbbsturl{https://inspirehep.net/record/863349/files/arXiv:1007.4931.pdf}}.

\bibitem{Beisert:2004hm}
\nbbstauthor{N.~Beisert, V.~Dippel and M.~Staudacher},
\nbbsttitle{``{A Novel long range spin chain and planar N=4 super
  Yang-Mills}''},
\nbbstjournal{\doiref{10.1088/1126-6708/2004/07/075}{JHEP~0407,~075~(2004)}},
\nbbsteprint{\arxivref{hep-th/0405001}{hep-th/0405001}}.

\bibitem{Basso:2013vsa}
\nbbstauthor{B.~Basso, A.~Sever and P.~Vieira},
\nbbsttitle{``{Spacetime and Flux Tube S-Matrices at Finite Coupling for N=4
  Supersymmetric Yang-Mills Theory}''},
\nbbstjournal{\doiref{10.1103/PhysRevLett.111.091602}{Phys.~Rev.~Lett.~111,~091602~(2013)}},
\nbbsteprint{\arxivref{1303.1396}{arxiv:1303.1396}}.

\bibitem{Eden:2016xvg}
\nbbstauthor{B.~Eden and A.~Sfondrini},
\nbbsttitle{``{Tessellating cushions: four-point functions in $\mathcal{N} $ =
  4 SYM}''},
\nbbstjournal{\doiref{10.1007/JHEP10(2017)098}{JHEP~1710,~098~(2017)}},
\nbbsteprint{\arxivref{1611.05436}{arxiv:1611.05436}}.

\bibitem{Fleury:2016ykk}
\nbbstauthor{T.~Fleury and S.~Komatsu},
\nbbsttitle{``{Hexagonalization of Correlation Functions}''},
\nbbstjournal{\doiref{10.1007/JHEP01(2017)130}{JHEP~1701,~130~(2017)}},
\nbbsteprint{\arxivref{1611.05577}{arxiv:1611.05577}}.

\bibitem{Bargheer:2009qu}
\nbbstauthor{T.~Bargheer, N.~Beisert, W.~Galleas, F.~Loebbert and
  T.~McLoughlin},
\nbbsttitle{``{Exacting N=4 Superconformal Symmetry}''},
\nbbstjournal{\doiref{10.1088/1126-6708/2009/11/056}{JHEP~0911,~056~(2009)}},
\nbbsteprint{\arxivref{0905.3738}{arxiv:0905.3738}}.

\bibitem{Borsato:2017lpf}
\nbbstauthor{R.~Borsato and A.~Torrielli},
\nbbsttitle{``{$q$-Poincar\'e supersymmetry in $AdS_5/CFT_4$}''},
\nbbsteprint{\arxivref{1706.10265}{arxiv:1706.10265}}.

\bibitem{Beisert:2017xqx}
\nbbstauthor{N.~Beisert, R.~Hecht and B.~Hoare},
\nbbsttitle{``{Maximally extended $\boldsymbol {\mathfrak{sl}(2\vert 2)}$ ,
  q-deformed $\boldsymbol{\mathfrak{d}{(2, 1;\epsilon)}}$ and 3D
  kappa-Poincar\'e}''},
\nbbstjournal{\doiref{10.1088/1751-8121/aa7a2f}{J.~Phys.~A50,~314003~(2017)}},
\nbbsteprint{\arxivref{1704.05093}{arxiv:1704.05093}}.

\bibitem{Gomez:2007zr}
\nbbstauthor{C.~Gomez and R.~Hernandez},
\nbbsttitle{``{Quantum deformed magnon kinematics}''},
\nbbstjournal{\doiref{10.1088/1126-6708/2007/03/108}{JHEP~0703,~108~(2007)}},
\nbbsteprint{\arxivref{hep-th/0701200}{hep-th/0701200}}.

\bibitem{Young:2007wd}
\nbbstauthor{C.~A.~S.~Young},
\nbbsttitle{``{q-deformed supersymmetry and dynamic magnon representations}''},
\nbbstjournal{\doiref{10.1088/1751-8113/40/30/033}{J.~Phys.~A40,~9165~(2007)}},
\nbbsteprint{\arxivref{0704.2069}{arxiv:0704.2069}}.

\bibitem{Costello:2013zra}
\nbbstauthor{K.~Costello},
\nbbsttitle{``{Supersymmetric gauge theory and the Yangian}''},
\nbbsteprint{\arxivref{1303.2632}{arxiv:1303.2632}}.

\bibitem{Costello:2017dso}
\nbbstauthor{K.~Costello, E.~Witten and M.~Yamazaki},
\nbbsttitle{``{Gauge Theory and Integrability, I}''},
\nbbsteprint{\arxivref{1709.09993}{arxiv:1709.09993}}.

\bibitem{Costello:2018gyb}
\nbbstauthor{K.~Costello, E.~Witten and M.~Yamazaki},
\nbbsttitle{``{Gauge Theory and Integrability, II}''},
\nbbsteprint{\arxivref{1802.01579}{arxiv:1802.01579}}.

\end{thebibliography}

\end{document}